\documentclass[preprint,aps,tightenlines,showpacs,nofootinbib,superscriptaddress]{revtex4}
\usepackage{color}
\usepackage{enumerate}
\usepackage{latexsym}
\usepackage[dvips]{graphicx}
\def\be{\begin{equation}}
\def\ee{\end{equation}}
\def\ba{\begin{eqnarray}}
\def\ea{\end{eqnarray}}
\newcommand{\beq}{\begin{eqnarray}}
\newcommand{\eeq}{\end{eqnarray}}
\newcommand{\eq}{eqnarray}

\newcommand{\al}{{\alpha}}
\newcommand{\ci}{\cite}

\newcommand{\de}{{\delta}}

\newcommand{\la}{{\lambda}}
\newcommand{\La}{{\Lambda}}

\newcommand{\ka}{{\kappa}}
\newcommand{\si}{{\sigma}}

\newcommand{\om}{{\omega}}

\newcommand{\no}{{\nonumber}}
\newcommand{\f}{\frac}

\newcommand{\ra}{\rightarrow}

\begin{document}

\preprint{arXiv:1808.03748v2 [hep-th]}
\today\\

\title{ 
Quasi-Normal Modes of a 
{\it Natural} AdS Wormhole 
in Einstein-Born-Infeld Gravity}

\author{Jin Young Kim \footnote{E-mail address: jykim@kunsan.ac.kr}}

\author{Chong Oh Lee \footnote{E-mail address: cohlee@gmail.com}}
\affiliation{Department of Physics, Kunsan National University,
Kunsan 54150, Korea}

\author{Mu-In Park \footnote{E-mail address: muinpark@gmail.com{, Corresponding author}
 }}
\affiliation{Research Institute for Basic Science, Sogang University,
Seoul, 121-742, Korea}

\begin{abstract}
We study the
matter perturbations of {a
new AdS wormhole} 
in (3+1)-dimensional Einstein-Born-Infeld gravity, called ``natural wormhole", which does not
require exotic matters.
{We discuss the 
stability of the perturbations by numerically computing
the quasi-normal modes (QNMs) of a massive scalar field in
the wormhole background.} We investigate
the dependence of quasi-normal frequencies on the mass of
scalar field 
as well as other
parameters of the wormhole.
It is found that the perturbations are always {stable}
for the wormhole geometry which has the general relativity {(GR)} limit
when the scalar {field mass} $m$ satisfies a
certain, tachyonic mass bound $m^2>m_*^2$ with $m_*^2<0$,
analogous to {the} Breitenlohner-Freedman (BF)
bound in the global-AdS space, $m_{\rm{BF}}^2=3 \La/4$. It is also found that the BF-like bound $m_*^2$ shifts by
the changes of the cosmological constant $\La$ or angular-momentum
number $l$, with a
{{\it level crossing}} between the lowest {complex
and pure-imaginary}
modes 
for zero {angular momentum}
$l=0$.
{Furthermore, it is found that the unstable modes can also
have oscillatory parts
as well as non-oscillatory parts
{\it depending on whether the real and imaginary parts of frequencies are
dependent on each other or not}, contrary to arguments in the literature.}
For wormhole geometries
{which do not have the} GR
limit, 
the BF-like bound does not occur and the perturbations
{are} stable
for arbitrary tachyonic and non-tachyonic masses{, up to a
critical mass $m_{c}^2>0$ where the perturbations are completely frozen.}

\end{abstract}

\pacs{04.20.Jb, 04.20.Dw, {04.62.+v, 04.70.Dy}, 11.10.Lm}

\maketitle

\newpage

\section{Introduction}

{Oscillations} in closed systems with conserved energies are described by normal
modes with real frequencies. Free fields 
propagating in {a confining box}
correspond to those cases.
On the other hand, in open systems with energy dissipations, oscillations are described by quasi-normal modes (QNMs) with complex frequencies.
Perturbations of fields propagating in the background of a black hole
correspond to these cases.
In the presence of a black hole, the (gravitational or matter) fields
can fall into the black hole so that
their energies can dissipate into the black holes and the fields decay

{The complex frequencies of QNMs carry the characteristic properties of {a black hole} like mass, charge, and angular momentum, {\it independent} of the initial perturbations.}
Because the computational details of QNMs depend much on the based gravity theories, QNMs can reflect the information about the gravity {theory itself} as well as that of {the
background spacetime}. In the recent detection of gravitational
waves by LIGO and {VIRGO, which are} thought to be radiated from mergers of binary black
holes or neutron stars,
 there is 
a regime, called {\it ring-down phase}, which
can be described by QNMs 
\cite{Abbo:2016,Abbo:2017}.
As more precise data
be available in the near future, QNMs
will be one of the key tools for the test of general relativity (GR) as well as {the black hole spacetime} \cite{Abbo:2016b}.

QNMs of black hole systems have been studied for a long time and much have
been known in various {cases}
(for some recent reviews, see \cite{Bert:2009}).
There is another important system, {\it wormhole} spacetime, which
{corresponds} to an open system so that QNMs appear too. Fields
entering into {a} wormhole without return would be {also} observed as the modes with
dissipating {energy and decay}. 

{Because the {characteristics of the} metric near the throat of a wormhole is different from those near the horizon of a black hole, one can distinguish them by comparing their QNMs of the gravitational waves with the same
boundary condition at asymptotic infinity.}
It has been found that wormhole geometries can also show the similar
gravitational wave forms as in the black hole systems up to some
early ring-down phase but some different wave forms at later times
\cite{Damo:2007,Card:2016} (see also \cite{Kono:2005} for relevant discussions).
In the near future, with increased precisions {of
gravitational wave detections}, it may be possible to
distinguish those two systems
{by investigating their late-time behaviors}. 

Compared to QNMs in the black hole systems, there are several issues about the wormholes themselves, which are still thought to be some hypothetical objects without any conclusive observational evidence, even though they can be exact solutions of Einstein's equation. In the conventional approaches to construct wormholes, there are the ``naturalness problems" due to
(i) the hypothetical exotic matters which support the throat of a wormhole but violate energy conditions \cite{Morr:1988} and (ii) the artificial construction of the wormhole throat by cuts and pastes \cite{Viss:1995}. Actually, in the recent analysis of gravitational waves from wormholes \cite{Damo:2007,Card:2016}, the considered wormholes are known as the ``thin-shell" wormholes which are quite
{artificial} \cite{Viss:1995}.

Recently a new type of wormhole solutions was proposed to avoid the problem caused by the exotic matters \cite{Cant:2010,SKim:2015}. In the new type of solutions, named as `` natural
wormholes" \cite{SKim:2015}, the throat is defined as the place where the solutions are {\it smoothly} joined. The metric and its
{derivatives} are continuous so that the exotic matters are not introduced at the throat.
From the new definition, throat can not be constructed arbitrarily contrary to the conventional cuts and pastes approach. The purpose of this paper is to study QNMs for these natural wormholes.

In this paper, we consider the recently constructed Anti-de Sitter (AdS)
wormholes in Einstein-Born-Infeld gravity \cite{JKim:2016}
and 
compute QNMs of a {\it massive} scalar field perturbation in the wormhole background. We consider {the} asymptotically AdS case since it is simpler than that of the asymptotically flat or de Sitter. Moreover there are several interesting aspects which are absent or unclear in other cases.

For example, it is well known that there exists a tachyonic mass bound, called Breitenlohner-Freedman (BF) bound, $m^2 > m_{\rm{BF}}^2$, for the ``conserved and positive" energy of perturbations of a massive scalar field with mass $m$ in the global AdS background \cite{Brei:1982} (see \cite{Gove:2008} for generalization to higher spins). The solutions exist with discrete real frequencies of ordinary normal modes above the BF bound and the perturbation becomes unstable below the BF bound. What we are going to {study} in this paper is about what happens in the BF bound for wormholes in the asymptotically AdS spacetime. It would be physically clear that the {\it local} deformation of a spacetime by the presence of wormholes would not change the stability properties of the whole spacetime much from the stability {property} at asymptotic infinity which {is} governed by the BF bound: It is hard to imagine a {\it smooth} spacetime where the (matter) perturbations with $m^2<m_{\rm{BF}}^2$ are {\it partly} unstable at infinity but also {\it partly} stable near wormholes, by some local effects. This implies that the perturbations of massive fields in {the background of AdS wormholes} would show both QNMs and BF bound. We will numerically compute these for a minimally-coupled real scalar field based on the approach of Horowitz and Hubeny for AdS space \cite{Horo:1999}. Moreover, the asymptotically AdS case would be interesting in the string theory contexts of the AdS/CFT correspondence. QNMs for AdS black hole spacetimes have been much studied in this context \cite{Wang:2000,Birm:2001} but little is known for AdS wormhole spacetimes.

The existence of large charged black holes
or wormholes would be questionable since our {universe} seems to be
charge-neutral in the large scales. However in the small scales, the
charged black holes or wormholes may exist, as the charged
elementary particles do. The well-known charged black hole is
Reissner-Nordstrom (RN) black hole with the usual Maxwell's
electromagnetic field in GR. But at short distances, we need
some modifications of GR for a consistent quantum theory, {\it i.e.},
(renormalizable) quantum gravity \cite{SKim:2015}. We may also
need {modification} of Maxwell's electromagnetism as
an effective description of quantum
effects or genuine classical modifications at short distances.
The non-linear generalization of Maxwell's theory by Born and Infeld (BI)
corresponds to the latter case \cite{Born:1934} and in this set up we may
consider the generalized charged black holes and wormholes which include
the RN case 
{as} the GR limit \cite{Garc:1984,Fern:2006}.
{On the other hand,} with the
{ advent} of D-branes, the BI-type action has {also} attracted renewed interests as an effective description of low energy superstring theory \cite{Leig:1989}.

The organization of the paper is as follow. In Sect. 2,
we 
consider the new AdS 
wormhole in Einstein-Born-Infeld (EBI)
gravity. In Sec. 3, we
set up a formalism 
to calculate QNMs of a massive scalar field.
In Sect. 4, we set up the formula for numerical computation of QNMs. {In Sec. 5,}
we summarize our numerical results of QNMs.
In Sec. 6, we conclude with some discussions. Throughout this paper, we use the conventional units for the speed of light $c$ and the Boltzman's constant $k_B$, $c=k_B=1$, but keep the Newton's constant $G$ and the Planck's constant $\hbar$ unless stated otherwise.

\section{New AdS wormholes in EBI gravity}
In this section, we describe a new AdS wormhole solution in EBI gravity which does not require exotic matters \cite{JKim:2016}. {To  this end}, we start by considering the EBI gravity action with a cosmological constant $\Lambda$ in $D=3+1$ dimensions,
\be
 S = \int d^{4} x \sqrt{-g} \left [ \frac{1}{16 \pi G}
 \left(R - 2\Lambda\right)
  + L(F)                            \right ],
 \label{EBI}
\ee
where $L(F)$ is the BI Lagrangian density, given by
 \be
 L(F) = 4 \beta^2 \left( 1 - \sqrt{ 1 + \frac{ F_{\mu\nu} F^{\mu\nu} }{2 \beta^2} }
                  \right) .
 \label{BI}
 \ee
Here, the parameter {$\beta~$} is a coupling constant with dimensions
$[{\rm length}]^{-2}$ which flows to infinity to recover the
usual Maxwell's electrodynamics
at low energies.

Taking $16 \pi G = 1$ for simplicity, the equations of motion are
obtained as
\be
\nabla_{\mu} \left ( \frac{F^{\mu\nu}}{\sqrt{ 1 + \frac{ F^2 }{2 \beta^2} }}
             \right ) = 0 ,
\label{eomforBI}
\ee
\be
 R_{\mu\nu} - \frac{1}{2}R g_{\mu\nu} + \Lambda g_{\mu \nu}=\f{1}{2} T_{\mu\nu},
 \label{eomforgrav}
 \ee
where the energy-momentum tensor for BI fields is given by
 \be
  T_{\mu\nu}=g_{\mu\nu} L(F) +\frac{4 F_{\rho\mu} F^\rho_{\nu} }{\sqrt{ 1 + \frac{ F^2 }{2 \beta^2} }} .
   \label{Tmunu}
   \ee
For the static and spherically symmetric metric ansatz,
 \be
 ds^2 = - N^2 (r) dt^2 + \f{1}{f(r)} dr^2 + r^2 (d \theta^2 + \sin^2 \theta d \phi^2),
 \ee
 and the electrically charged case where the only non-vanishing component of the field-strength tensor is $F_{rt} \equiv E(r) $, the general solution is given by
 \beq
 N^2(r)
&=&f(r)=
1 - \frac{2 M}{r} - \frac{\Lambda}{3} r^2
 + \frac{2 }{3} \beta^2 r^2 \left ( 1 - \sqrt{ 1+ \frac{Q^2} { \beta^2 r^{4} } } \right )
  + \frac{4}{3} \frac{Q^2} {r^{2}}~ {_2 F}_1 \left ( \frac{1}{2} , \frac{1}{4}; \frac{5}{4};
  \frac{-Q^2}{\beta^2 r^{4}} \right ),
   \label{f:sol2} \\
   E(r) &=& \frac{Q}{\sqrt{ r^{4} + \frac{Q^2}{\beta^2}    } } ,
 \label{Esol}
\eeq
in terms of the hypergeometric function \cite{Garc:1984}. Here $Q$ represents the electric charge located at the origin and $M$ is the ADM mass which is composed of the intrinsic mass $C$ and (finite) self energy of a point charge $M_0$, defined by
\beq
M &= &C+ M_{0}, \\
M_{0} &=& \frac{2}{3} \sqrt{ \frac{\beta Q^3}{\pi} } \Gamma\left( \frac{1}{4} \right) \Gamma\left( \frac{5}{4} \right).
\label{M0}
 \eeq
The metric function has the different behavior depending on $\beta Q$ and ADM mass $M$ (Fig. 1).

In the construction of {\it natural} wormholes, the throat which connects two
universes (or equivalently, two remote parts of the same universe) is
defined as the place where the solutions are smoothly joined. For the
reflection $(Z_2)$ symmetric universe, the new
spherically symmetric wormhole metric is described by
\begin{\eq}
  ds^2=-N_{\pm}(r)^2 dt^2+\frac{dr^2}{f_{\pm}(r)}+r^2
\left(d\theta^2+\sin^2\theta d\phi^2\right)
\label{wormhole}
\end{\eq}
when there exits the throat $r_0$, defined by
\begin{\eq}
\left.\f{dN_{\pm}}{dr}\right|_{r_0}=\left.\f{df_{\pm}}{dr}\right|_{r_0} =0,
\label{throat}
\end{\eq}
and the matching condition,
\begin{\eq}
N_+(r_0)=N_-(r_0), ~f_+(r_0)=f_-(r_0),
\label{metric_contin}
\end{\eq}
with two coordinate patches, each one covering the range $[r_0, +\infty)$.
If there is a singularity-free coordinate patch ${\cal M}_+$ for all values of
$r \geq r_0$, one can construct a smooth regular wormhole-like geometry,
by joining ${\cal M}_+$ and its mirror patch ${\cal M}_-$ at the throat $r_0$.

Note that, in this new definition, throats can not be constructed arbitrarily contrary to the conventional cuts and pastes approach. Moreover, in the new approach, $f_\pm(r_0)$ needs {\it not} to be vanished
in contrast to {Morris-Thorne's} approach \ci{Morr:1988}, while the quantities ${dN_{\pm}(r_0)}/{dr},~ {df_{\pm}(r_0)}/{dr}$ in (\ref{throat}) need {\it not} to be vanished in both {Morris-Thorne's approach \ci{Morr:1988} and  Visser's} cuts and pastes approach  \ci{Viss:1995}.

\begin{figure}
\includegraphics[width=4.8cm,keepaspectratio]{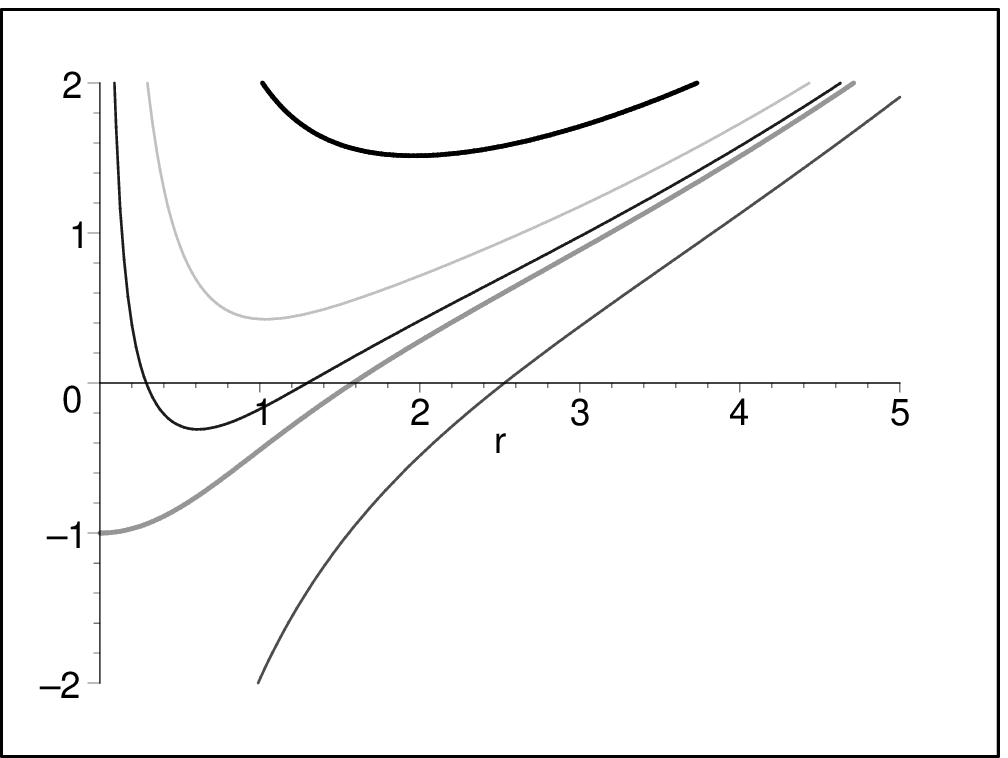}
\qquad
\includegraphics[width=4.8cm,keepaspectratio]{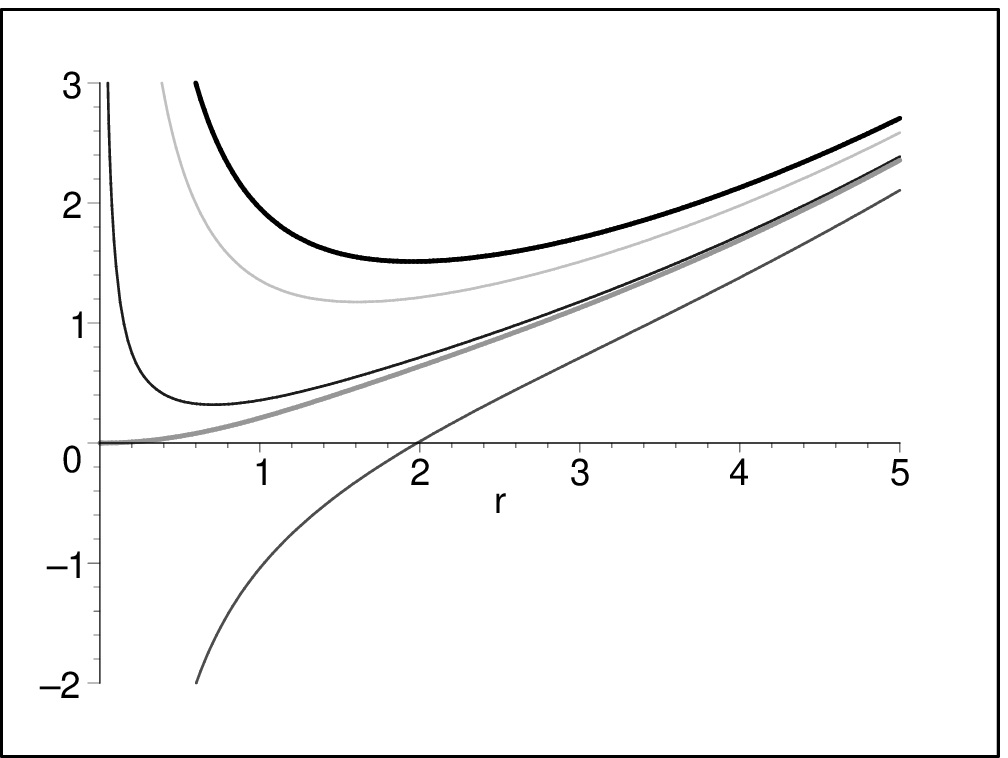}
\qquad
\includegraphics[width=4.8cm,keepaspectratio]{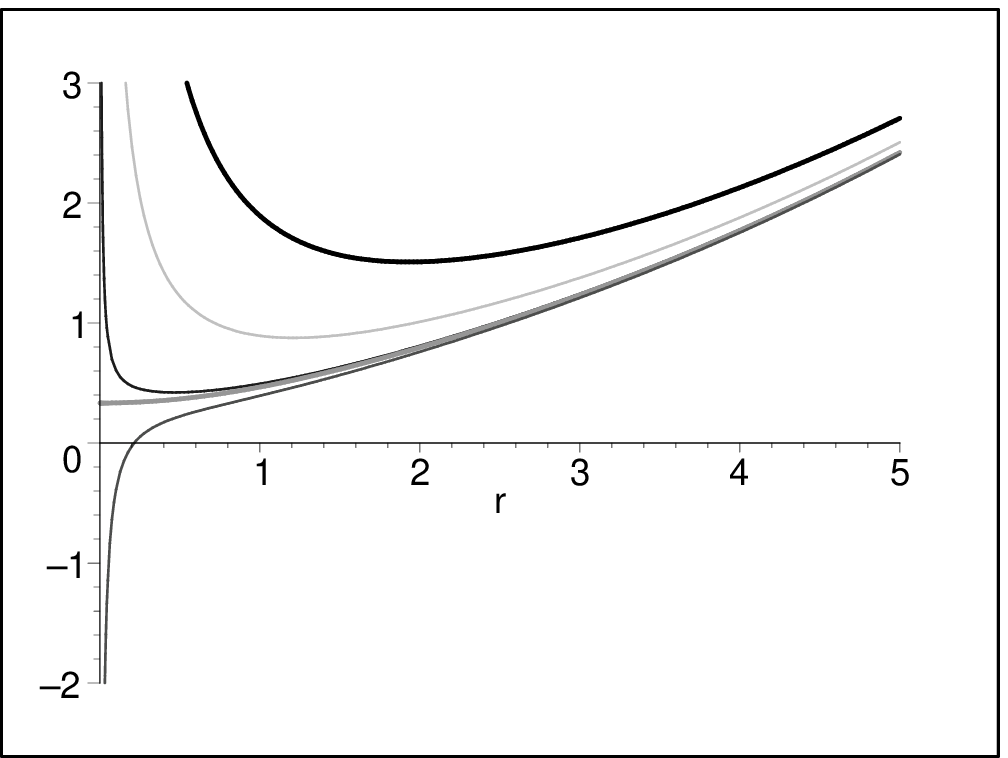}
\caption{The plots of $f(r)$ for {
varying}
$M$ with a fixed $\beta Q$ ($\beta Q >1/2$ (left),
$\beta Q=1/2$ (center), $\beta Q<1/2$ (right)) and a
negative cosmological constant $\Lambda<0$ . We consider
{(top to bottom)} $M=0,~0.95,~1.2,~M_0,~1.5$ with
{ $\beta=1$, $M_0\approx 1.236$ (left);  $M=0,~0.85,~M_0,~0.9,~1.2$
with $\beta=1/2$, $M_0 \approx 0.874$ (center); and $M=0,~0.7,~M_0,~0.75,~1.2$
with $\beta=1/3$, $M_0 \approx 0.714$ (right)}, respectively for $Q=1, \Lambda=-1/5$.}
\label{fig:f}
\end{figure}

\begin{figure}
\includegraphics[width=7cm,keepaspectratio]{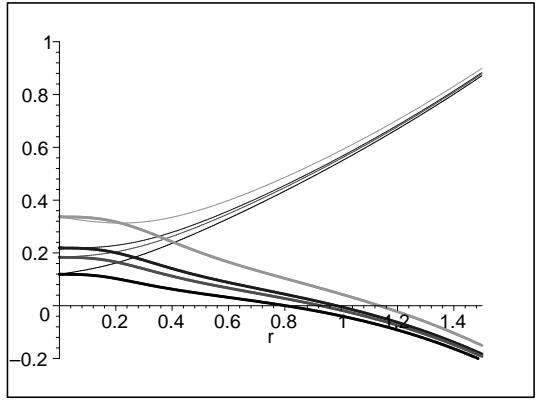}
\qquad
\caption{The plots of the ADM mass $M$ vs. the wormhole radius $r_0$ (thick curves) and the black hole horizon radius $r_+$ (thin curves) for
{
varying}
$\beta Q$ with a fixed cosmological constant $\Lambda<0$. The marginal mass
$M_0$ is given by the mass value at $r_0$ or $r_+=0$. {For wormholes,
$M(r_0)$ is a monotonically decreasing function with the maximum value $M_0$
at $r_0=0$. For black holes, $M(r_+)$ is a monotonically increasing function with the
minimum value $M_0$ at $r_+=0$ when $\beta Q \leq 1/2$ (three thin curves from
below) and concave with
{the} minimum $M^*<M_0$ at the extremal horizon $r_+^*$ when
$\beta Q>1/2$ (top thin curve).}
We consider $\beta Q={2/3,~1/2,~2/4.5,~1/3}$ (top to bottom) with $\beta=2$ and $\Lambda=-1/5$. }
\label{fig:M}
\end{figure}

In Fig.1 for the solution (\ref{f:sol2}) one can easily see the existence of the throat $r_0$ satisfying the conditions (\ref{throat}) and (\ref{metric_contin}), depending on the mass $M$ for given values of $\beta, Q,$ and $\La$. Now, from the property of the metric function \cite{Garc:1984}
 \be
 \frac{d}{dr} (r f ) =f+r \f{df}{dr}=  1 - \Lambda r^2
 + 2 \beta^2 r^2 \left (1  - \sqrt{ 1 + \frac{ Q^2} { \beta^4 r^{4}}  } \right ),
 \label{f:Eq}
 \ee
one can find that at the throat $r_0$, the largest $r$ of $df/dr =0$,
\be
 f(r_0)=  1 - \Lambda r_0^2
 + 2 \beta^2 r_0^2 \left ( 1 - \sqrt{ 1 + \frac{ Q^2} { \beta^2 r_0^{4}}  } \right ).
 \label{f:Eq_r0}
 \ee
Comparing (\ref{f:Eq_r0}) with the general solution (\ref{f:sol2}), the { wormhole mass} $M$ can be expressed in terms of $r_0$ as,
\be
 M(r_0)= \f{r_0^3}{3} \left[
 \Lambda
  -2 \beta^2 \left ( 1 - \sqrt{ 1+ \frac{Q^2} { \beta^2 r_0^{4} } } \right )
  + \frac{2 Q^2} {r_0^4}~ {_2 F}_1 \left ( \frac{1}{2} , \frac{1}{4}; \frac{5}{4};
  \frac{-Q^2}{\beta^2 r_0^{4} } \right ) \right],
  \label{M_r0}
\ee
which is a {monotonically} decreasing function of $r_0$ with the {\it maximum} value
$M_0$ of (\ref{M0}) at $r_0=0$ ({thick} curves in Fig. 2).
It {is interesting to note}
that the mass of our new AdS wormhole without exotic matters
can be {\it negative} for large $r_0$, similar to the conventional wormholes which require the exotic matters and violate energy conditions. This may be considered as another evidence that {\it exotic matters cemented at the throat are not mandatory for constructing large scale wormholes.}

\begin{figure}
\includegraphics[width=7cm,keepaspectratio]{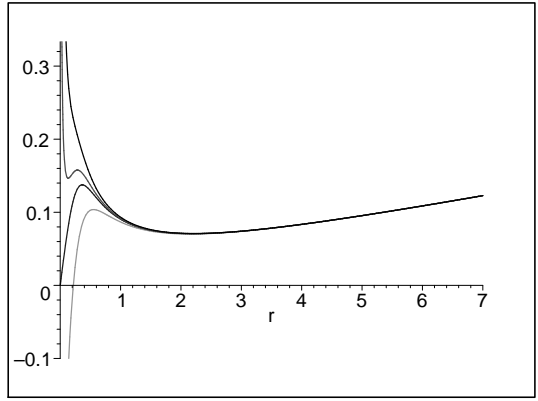}
\caption{The plots of the Hawking temperature $T_H$ {or $f(r_0)/r_0$} vs. the black hole horizon radius $r_+$ {or the wormhole throat radius $r_0$} for
{
varying}
$\beta Q$ with a fixed cosmological constant $\Lambda<0$. We consider $\beta Q={2/3,~1/2,~2/4.5,~1/3}$ (bottom to top curves) with $\beta=2$, $\Lambda=-1/5$.}
\label{fig:T}
\end{figure}

{(\ref{M_r0})} is in contrast to the black hole mass $M$ in terms of the black hole horizon $r_+$, corresponding to the largest $r$ of $f(r)=0$,
\beq
M(r_+)=\f{r_+}{2} \left[ 1 -  \frac{\Lambda}{3} r_+^2
 + \frac{2 }{3} \beta^2 r_+^2 \left ( 1 - \sqrt{ 1+ \frac{Q^2} { \beta^2 r_+^{4} } } \right )
  + \frac{4}{3} \frac{Q^2} {r_+^{2}}~ {_2 F}_1 \left ( \frac{1}{2} , \frac{1}{4}; \frac{5}{4};
  \frac{-Q^2}{\beta^2 r_+^{4}} \right ) \right],
  \label{M_bh}
\eeq
which is
{{\it positive}} definite. This is a {monotonically}
increasing function of $r_+$ with the {\it minimum} $M_0$ at $r_+=0$ when
$\beta Q \leq 1/2$ so that there exists only {the Schwarzschild-like (type I)}
black hole with {one} horizon (three thin curves from below in Fig. 2).
On the other hand, when $\beta Q > 1/2$ (top thin curve in Fig. 2), there
exists {the RN-like (type II) black hole} with two horizons as well as
{the Schwarzschild-like}
black hole with one horizon. {In this latter case,} the black hole mass
function (\ref{M_bh}) {is concave 
with the} minimum
\beq
M^*=\f{r_+^*}{3} \left[ 1
  + \frac{2 Q^2} {{r_+^*}^{2}}~ {_2 F}_1 \left ( \frac{1}{2} , \frac{1}{4}; \frac{5}{4};
  \frac{-Q^2}{\beta^2 {r_+^*}^{4}} \right ) \right],
\eeq
{which is smaller than $M_0$}, at the ``extremal" horizon $r_+^*$, where the outer horizon $r_+$ meets the inner horizon $r_-$ at
\beq
r^*_+=\sqrt{
\f{\La-2 \beta^2
+ 2 \sqrt{(\beta^2-\La/2)^2+\La(\La-4 \beta^2)(\beta^2 Q^2-1/4)}
}
{\La(\La-4 \beta^2)}
}
\label{r:ext}
\eeq
with the vanishing Hawking temperature for the outer horizon $r_+$ (Fig.3),
\beq
T_H &\equiv& \f{\hbar}{4 \pi} \left. \f{df}{dr} \right|_{r=r_+} \no \\
&=&\f{\hbar}{4 \pi} \left[ \f{1}{r_+}-\La r_+ +2 \beta^2 r_+ \left(1-\sqrt{1+\f{Q^2}{\beta r_+^4}} \right) \right].
\eeq
{In the latter case, the {wormhole
mass $M$ increases as the throat radius $r_0$ reduces,} by accretion of ordinary (positive energy)
matters until it reaches to the extremal black hole horizon
$r_+^*$ where the wormhole mass is equal to the black hole mass {$M(r_0)=M(r_+)=M^*$}.
Then, no }further causal contact with the wormhole is possible ``classically" afterwards
since the throat {is located}
inside the horizon $r_+$\footnote{This implies that natural wormholes can be the {factories} of black holes by accretion of ordinary matters or vice versa \cite{SKim:2015}.}.
Hence, for {the case} $\beta Q>1/2$, the
ranges of $r_0$ and the wormhole mass $M(r_0)$ {are} bounded by
{$r_0 > r_+^*,~M(r_0) < M^*$} for the ``observable" wormhole throat
outside the black hole horizon $r_+$, in contrast to $r_0 \geq 0,
~M(r_0) \leq M_0$ for the former {case $\beta Q
\leq 1/2$.}

Finally we note that, near $r_0$, the metric function can be expanded as
\be
f(r)=f(r_0)+\f{\si}{2} (r-r_0)^2 +{\cal O}[(r-r_0)^3]
\label{f:near throat}
\ee
with the second moments,
\be
\si {\equiv} \left.\f{d^2f}{dr^2}\right|_{r_0}=-2 \La+ 4 \beta^2 \left[1-
\f{1}{r_0^{4}} \left(1+\f{Q^2}{\beta^2 r_0^4}\right)^{-3/2}
\right].
\label{sigma}
\ee
At large $r$, on the other hand, the metric function is expanded as
\be
 f(r) = 1 - \frac{\Lambda}{3} r^2 - \frac{2 M }{r}
 +  \frac{Q^2}{r^{2}} -\f{Q^4}{20 \beta^2 r^6} + {\cal O}(r^{-10}).
\label{f:larger}
 \ee

\section{Massive scalar perturbations in the new AdS-EBI wormhole}

In this section, we consider the perturbations of {a massive scalar field and its QNMs} in the new AdS-EBI wormhole background.
The wave equation for {a minimally-coupled} massive scalar field $\Phi(t,{\bf r})$ with mass $m$ is given by
\be
\frac{1}{\sqrt{-g}}\partial_{\mu}
\left(\sqrt{-g}g^{\mu\nu}\partial_{\nu}\Phi \right)-m^2\Phi=0.
\label{KGeq}
\ee
Considering the mode solutions
\be
\Phi(t,{\bf r})=e^{-i\omega t}Y(\theta,\phi)\frac{\widetilde{\varphi}(r)}{r},
\ee
with the spherical harmonics $Y(\theta,\phi)$ on ${\rm S}^2$, the wave equation reduces to the standard radial equation,
\be
\left(\frac{d^2}{dr_{*}^2}+\omega^2\right)\widetilde{\varphi}(r)= \widetilde{V}(r)\widetilde{\varphi}(r),
\label{Rad_Eq_tort}
\ee
where $r_{*}$ is the tortoise coordinate, defined by
\be
dr_*\equiv \frac{dr}{f(r)},
\ee
and $\widetilde{V}(r)$ is the effective potential, given by
\be
\widetilde{V}(r)=f(r)\left(\frac{l(l+1)}{r^2}+\frac{1}{r} \f{d f(r)}{dr}+m^2\right)
\label{V_tilde}
\ee
with the angular-momentum number $l$.

Choosing the tortoise coordinate $r_*=0$ at the throat $r_0$, one obtains
\beq
r_* &\equiv& \int^r_{r_0} f^{-1}(r) dr \no \\
&=&f^{-1}(r_0) (r-r_0) -\f{\si}{3 !} f^{-2}(r_0) (r-r_0)^3+{\cal O}[(r-r_0)^4]
\label{tort_throat}
\eeq
near the throat $r\approx r_0$ and
\be
r_*=\int^\infty_{r_0} f^{-1}dr+\f{3}{\La r}+\f{3}{\La^2 r^3}+{\cal O}(r^{-4})
\label{tort_infty}
\ee
at large $r$, using the asymptotic expansions (\ref{f:near throat}) and (\ref{f:larger}){, respectively}. Here, $\int^\infty_{r_0}f^{-1} dr$ is the (finite) value of the tortoise coordinate $r_*(r)$ evaluated at $r=\infty$ and can be expanded as $-{3}{\La^{-1}} (r_0^{-1}+{\La^{-1}} r_0^{-3}+ \cdots)$.

\begin{figure}
\includegraphics[width=7cm,keepaspectratio]{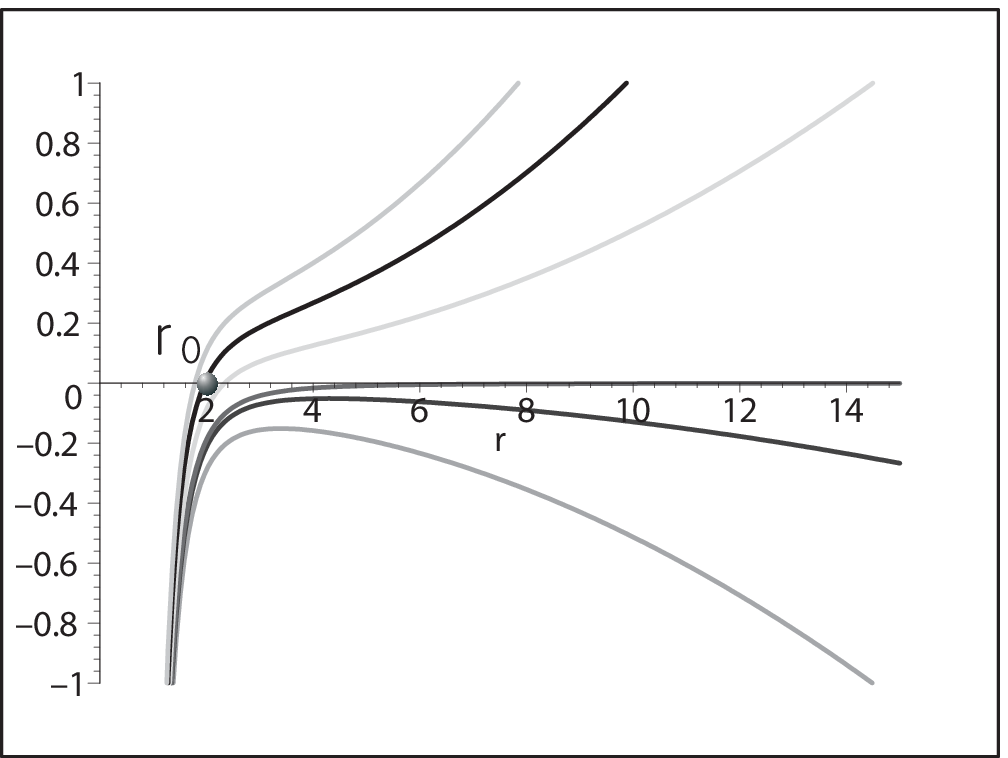}
\qquad
\includegraphics[width=7cm,keepaspectratio]{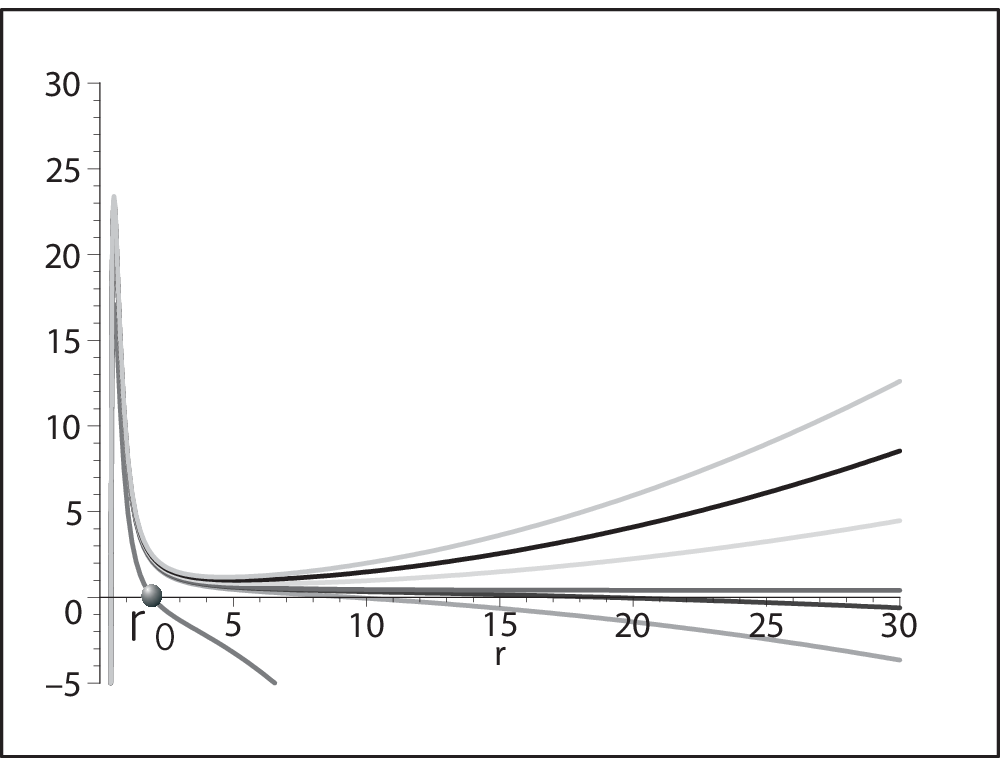}
\caption{The plots of the effective potential $\widetilde{V}(r)$ for
{
varying}
$m$ with a fixed $\beta Q>1/2$ and a cosmological constant $\La<0$. We consider $m^2=-\La/3,0,\La/3,2\La/3,3\La/4,\La,-l(l+1)/r_0^2$ (top to bottom curves) {for $l=0$ (left) and $l=2$ (right)} with $M=10^{-5},\La=-1/5,r_0 \approx 1.96314$. $\widetilde{V}(r)$ vanishes at the throat $r_0$ only for the mass satisfying $m^2=-l(l+1)/r_0^2${; $m^2=0$} for $l=0$ (left, second curve from top) { or} a tachyonic mass $m^2=-6/r_0^2$ for $l=2$ (right, bottom curve).}
\label{fig:tilde_V}
\end{figure}

From the near-throat behavior of the effective potential $\widetilde{V}$,
\be
\widetilde{V}(r)=f(r_0)\left(\f{l(l+1)}{r_0^2}+m^2\right)
+\f{f^2(r_0)}{r_0} \left(-\f{2l(l+1)}{r_0^2}+\si\right)r_*+\cdots,
\ee
the 
mode {solution near the throat is} obtained as
\be
\Phi (t,{\bf r}) \approx A_1 (\theta,\phi) \f{e^{-i(\om t+k r_*)}}{r}+A_2 (\theta,\phi) \f{e^{-i(\om t-k r_*)}}{r}
\label{Phi_near_throat}
\ee
with
\be
k=\sqrt{\om^2-\widetilde{V}(r_0)},
\label{k_throat_general}
\ee
{where $\widetilde{V}(r_0)=f(r_0) [{l(l+1)}/{r_0^2}+m^2 ]$.} {Here, $A_1$ and $A_2$ parts
represent purely ingoing and outgoing modes, respectively.} Since
QNMs are defined as solutions which are purely ingoing near the throat,
we set $A_2(\theta,\phi) =0$ as our desired boundary condition at the throat. Here,
it is interesting to note that the solutions { 
at the throat} are {\it not light-like} ``generally" due to {\it non-vanishing} $f(r_0)$ { and the effective potential
$\widetilde{V}(r_0)$,}
in contrast to the {\it always-light-like} solutions {at the black hole horizon} $r_+$, where $f(r_+)$ and the effective potential (\ref{V_tilde}) vanish:
{The effective potential $\widetilde{V}(r_0)$ {\it may}
vanish}
when
the mass of scalar field $m$ satisfies $m^2=-l (l+1)/r_0^2$ (for example, $m=0$ for $l=0$, or $m^2=-6/r_0^2$ for $l=2$),
but generally it does not
\footnote{(i) $l=0$ modes: For $m=0$, $\widetilde{V}$ is
positive definite for $r > r_0$ and vanishes at the throat $r_0$, for $m^2>0$,
$\widetilde{V}$ is {positive} definite for the whole region of $r \geq r_0$,
whereas for $m^2<0$, $\widetilde{V}$ is not {positive} definite but it depends
on $m^2$. (ii) $l>0$ modes: For $m^2=-l (l+1)/r_0^2$, $\widetilde{V}$ vanishes
at $r_0$ but {it is} negative for $r > r_0$, whereas for $m^2 \geq 2 \La/3$,
$\widetilde{V}$ is positive for the whole region of $r \geq r_0$, otherwise,
 {\it i.e.,} for $-l (l+1)/r_0^2<m^2 <2 \La/3$, $\widetilde{V}$ is not {positive}
 definite. It is interesting to note that in the last {case}
 $\widetilde{V}(r_0)<0$, the ingoing waves become {``tachyonic" at the throat, {\it i.e.}, $k^2>\om^2$} and we will see later that these perturbations are still stable if it is not too much tachyonic, {\it i.e.,} $2 \La/3 > m^2 > 3 \La/4$.}  (Fig. 4).

On the other hand, at $r=\infty$, which corresponds to a finite value of
$r_*|_{r=\infty}=\int^{\infty}_{r_0} f^{-1}dr$, the effective potential
$\widetilde{V}(r)$ diverges as 
\be
\widetilde{V}(r) =-\f{\La}{3} r^2 \left( m^2-\f{2}{3} \La \right)
+ \left( m^2 -\f{2}{3} \La -\f{\La l(l+1) }{3} \right)
+{{\cal O}(r^{-1})}.
\ee
In terms of the tortoise coordinate $r_*$, we have
\beq
\widetilde{V}(r) &=&-\f{\La}{3} \left( m^2-\f{2}{3} \La \right)
\left(r_*- \int^{\infty}_{r_0} f^{-1}dr \right)^{-2}
+ \left[ \f{1}{3}\left(m^2 -\f{2}{3} \La \right)-\f{\La  l(l+1) }{3} \right] \no \\
&+&{{\cal O}\left(r_*- \int^{\infty}_{r_0} f^{-1}dr \right)},
\eeq
using $r=3 \La^{-1} \left(r_*- \int^{\infty}_{r_0} f^{-1}dr \right)^{-1}+\cdots$ for the  limit $r\ra \infty$, from (\ref{tort_infty}). Then, the radial wave equation (\ref{Rad_Eq_tort}) reduces to
\be
\left[\frac{d^2}{d\widetilde{r}_{*}^2}+\omega^2+\f{\La }{3}\left( l(l+1)-\f{\la}{3} \right) +\f{\la}{\widetilde{r}_{*}^2}
\right]\widetilde{\varphi}(r)= {{\cal O}(\widetilde{r}_{*})},
\label{Rad_Eq_tort_infty}
\ee
where $\la \equiv 3 \La^{-1} ( m^2 -2\La/3 )$ and  $\widetilde{r}_{*} $ is the {\it shifted} tortoise coordinate $\widetilde{r}_{*} \equiv r_*- \int^{\infty}_{r_0} f^{-1}dr$, which
 approaches zero as $r \ra \infty$.

Now, near $\widetilde{r}_{*}= 0$ ($r=\infty$), the leading order solution of (\ref{Rad_Eq_tort_infty}) is obtained as
\be
\widetilde{\varphi}(r) \approx B_1 \widetilde{r}_*^{\f{1}{2}(1+\sqrt{1-4 \la})}
+B_2 \widetilde{r}_*^{\f{1}{2}(1-\sqrt{1-4 \la})}.
\label{sol_tort_infy_tilde}
\ee
Since the norm of the wave function $\Phi$ is given by
\be
\int dx^3 \Phi^* \Phi
\sim \int^{\infty}_{r_0} dr \widetilde{\varphi}^* \widetilde{\varphi}
\sim \int^{0}_{-\delta} \f{d\widetilde{r}_*}{\widetilde{r}_*^2} \widetilde{\varphi}^* \widetilde{\varphi},
\ee
where $\de \equiv \int^{\infty}_{r_0} f^{-1}dr~(<\infty)$, the solution (\ref{sol_tort_infy_tilde}) is square-integrable only for $B_2=0$ and
\be
\la < \f{1}{4},
\ee
or equivalently \footnote{For $\la = 1/4$, there is a logarithmic divergence at $\widetilde{r}_*=0$ and the {solution is} not normalizable, in contrast to purely lower-dimensional problems \cite{Calo:1969}. (cf. \cite{Case:1950})}
\be
m^2 >\f{3 \La}{4}.
\label{BF_bound}
\ee
This result agrees also with the {condition of} regularity or finite energy of the {solution} \cite{Calo:1969,Brei:1982}. In particular, (\ref{BF_bound}) represents the stability condition of massive scalar perturbations in the {\it global} AdS space \footnote{In some literature (cf. \cite{Brei:1982}), the limiting case $m^2=3 \La/4$ {, or more generally in $D$ dimensions, $m^2=(D-1) \La/2(D-2)$ or $-(D-1)^2 \ell^{-2}/4$ with $\ell^{-2}=-2 \La/(D-1)(D-2)$,} has been classified as the stable one due to positivity of {the} energy functional but it would not be a physically viable fluctuation due to {the divergence of its} energy functional, which is related to {the divergent norm of the solution} as discussed in the above footnote No. $3$.}, {$m^2>m^2_{\rm{BF}}$ with the BF bound $m^2_{\rm{BF}}=3 \La/4$} \cite{Brei:1982,Moro:2010}. Moreover, note that the solution (\ref{sol_tort_infy_tilde}) with the bound (\ref{BF_bound}), satisfies the vanishing Dirichlet boundary condition at $r=\infty$ ($\widetilde{r}_*=0$) even though the effective potential $\widetilde{V}$ is not {positive} infinite {(Fig. 4): At} $r=\infty$, $\widetilde{V}$ is {positive} infinite for $m^2 >2 \La/3$ but zero or {negative} infinite for $2 \La/3 \geq m^2 > 3 \La/4$. The usual stability criterion based on the positivity of the effective potential is not quite correct when considering massive perturbations in AdS background for the latter mass range \footnote{The bound $m^2 > 3 \La/4$ corresponds to the absence of ``genuine" tachyonic modes in {the global} AdS background \cite{Brei:1982}. But this does not mean the absence of {genuine tachyonic modes} {\it locally}. Actually, for $2 \La/3 > m^2 > 3 \La/4$, the ingoing modes at the throat are tachyonic as noted in the footnote No. 2, in contrast to QNMs in black hole background which are always light-like at the black hole horizon. } \cite{Chan:1985}.

\section{{\it Massive}~ Quasi-Normal Modes}

QNMs are defined as the solutions which are purely ingoing near the throat
$\Phi \sim e^{-i(\om t+k r_*)}$. In this paper, {our interest is}
the dependence of QNMs on the mass of perturbed fields and in this section we will consider their computations, which are
{called}
``massive" QNMs \cite{Ohas:2004}, based on the approach of Horowitz and Hubeny (HH) \cite{Horo:1999}. In order to study QNMs, it is convenient to work with the Eddington-Finkelstein coordinate by introducing the ingoing null coordinate $v=t+r_*$ with the metric
\be
ds^2=-f(r)dv^2+2dvdr+r^2(d\theta^2+\sin^2\theta d\phi^2).
\label{EF_metric}
\ee
Considering the mode solution,
\be
\Phi(t, {\bf r})=e^{-i\omega v}Y(\theta,\phi)\frac{\varphi(r)}{r},
\ee
the wave equation (\ref{KGeq}) reduces to the radial equation for $\varphi(r)$,
\be
f(r)\frac{d^2\varphi(r)}{dr^2}+\left(\f{d f(r)}{dr}-2 i \omega\right)\frac{d\varphi(r)}{dr}-V(r)\varphi(r)=0
\label{Rad_Eq_EF_coord}
\ee
with the {\it reduced} effective potential $V(r)$,
\be
V(r)=\frac{l(l+1)}{r^2}+\f{1}{r}\frac{d f(r)}{dr}+m^2.
\ee
Note that, as in the effective potential $\widetilde{V}(r)$, the reduced
potential $V(r)$ is {\it not} {positive} definite and its positivity depends
on $m^2$ (Fig. 5): $V(r)$ is {positive} definite for the whole region of
$r \geq r_0$ only for $m^2> 2 \La/3$.

\begin{figure}
\includegraphics[width=6cm,keepaspectratio]{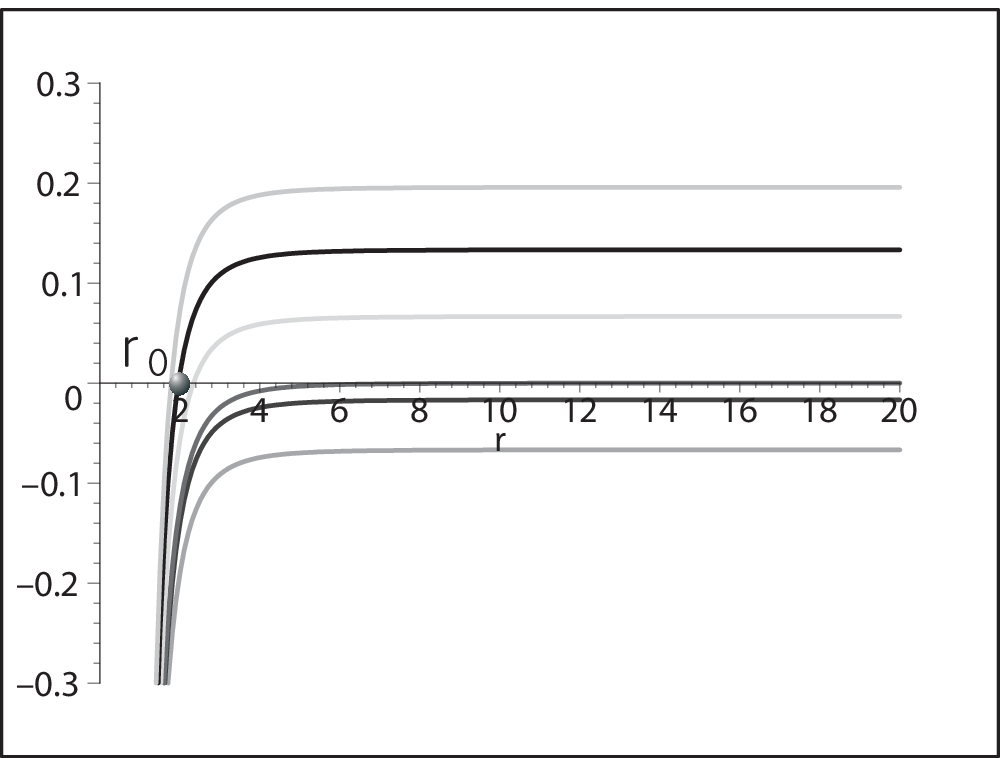}
\qquad
\includegraphics[width=6cm,keepaspectratio]{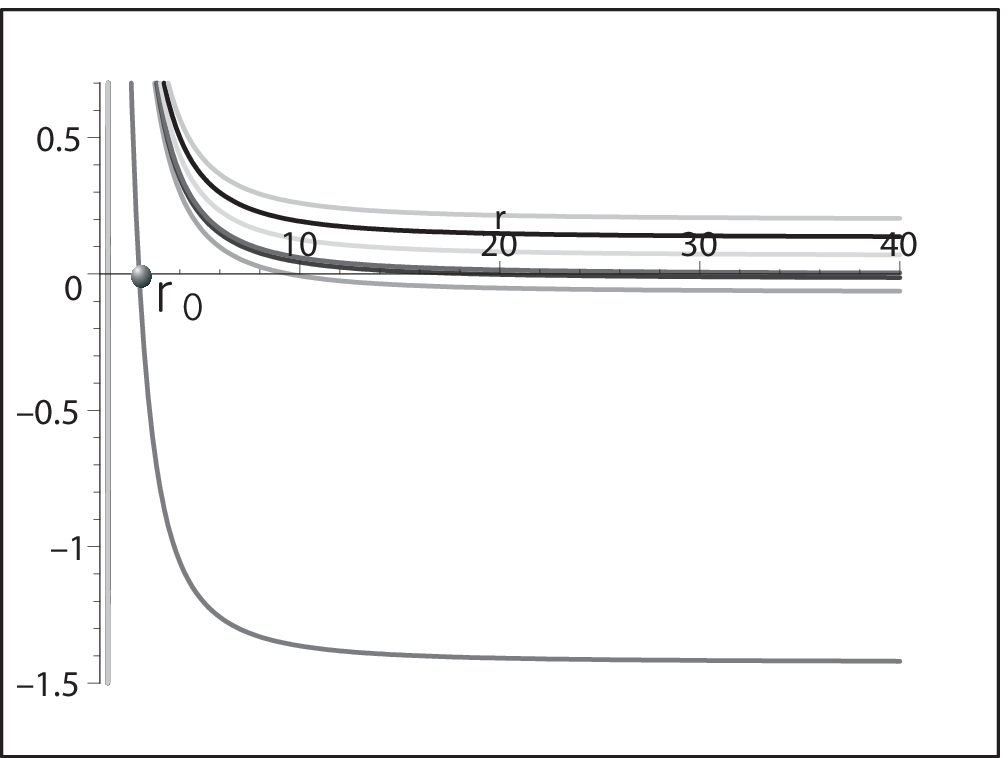}
\caption{The plots of the {\it reduced} effective potential ${V}(r)$ for the corresponding {effective} potential $\widetilde{V}(r)$ in Fig. 4.}
\label{fig:V}
\end{figure}

To compute QNMs, we will expand the solution {as} a power series around the wormhole throat $r_0$ and impose the vanishing Dirichlet boundary conditions at $r=\infty$, following the approach of HH \cite{Horo:1999}. In order to treat the whole region of interest, $r_0 \leq r <\infty$, into a finite region, we introduce a new variable $x=1/r$ so that the metric (\ref{EF_metric}) becomes \footnote{{The metric approaches to that of $AdS_4$ in Poincare patch $ds^2=(dz^2+\eta^{\mu \nu}dx_{\mu} dx_{\nu})/z^2$ with the three-dimensional flat Minkowski metric $\eta^{\mu \nu}$ and the radial {\it AdS} coordinate $z\sim x$ for an appropriate choice of scalings. { (cf. \cite{Moro:2010})}}}
\be
ds^2=-f(x)dv^2+\f{1}{x^2}(-2dvdx+d\theta^2+\sin^2\theta d\phi^2).
\label{EF_metric_x}
\ee
The scalar field equation can be written as
\be
s(x)\frac{d^2\varphi(x)}{dx^2}
+\frac{t(x)}{x-x_0}\frac{d\varphi(x)}{dx}+\frac{u(x)}{(x-x_0)^2}\varphi(x)=0,
\label{Rad_Eq_EF_x}
\ee
where $x_0 = 1 /r_0$ and the coefficient functions are given by
\beq
s(x)&=&-\f{x^4 f(x)}{x-x_0}, \no \\
t(x)&=&-2 x^3 f(x)-x^4 \dot{f}(x)-2 i \om x^2, \no \\
u(x)&=&(x-x_0) \left[ l(l+1) x^2-x^3 \dot{f}(x)+m^2 \right].
\eeq
The overdot ($\dot{~}$) represents the derivative with respects to $x$.
Since $f(x_0)>0$ in our wormhole system, one can remove the overall $(x-x_0)^{-1}$ factor in (\ref{Rad_Eq_EF_x}) so that $x=x_0$ is not a singular point, whereas there is one regular singular point at the spatial infinity $x=0$.  {Then one} can expand equation (\ref{Rad_Eq_EF_x}) around the throat $x_0$ up to the pole at $x=0$ and solve the equation at each order of {the} expansion.

First, expanding $s,t,u$ around $x=x_0$ as
\beq
s(x) &=& \sum_{n=-1}^{\infty} s_n (x-x_0)^n,  \no \\
t(x) &=& \sum_{n=0}^{\infty} t_n (x-x_0)^n,     \no \\
u(x) &=& \sum_{n=0}^{\infty} u_n (x-x_0)^n,
\label{stu_expansion}
\eeq
one can obtain {the first few} coefficients of them as follows,
\beq
&&s_{-1} = - x_0^4 f(x_0),~s_{0} = -4 x_0^3 f(x_0),~s_{1} = - 6 x_0^2 f(x_0)-\f{1}{2}x_0^4 \ddot{f}(x_0), \no \\
&&t_0=-2 x_0^2 \left[ x_0 f(x_0)+i \om \right],~t_1=-x_0 \left[6 x_0 f(x_0)+x_0^3 \ddot{f}(x_0)+4i \om \right], \no \\
&&t_2=-2i \om+ ~{({\rm non-}\om~ {\rm terms})}~, ~
t_{n(>2)}= ~{({\rm non-}\om~ {\rm  terms})}~ ,\no \\
&&u_0=0,~u_1 = l(l+1) x_0^2 + m^2, ~u_2 =x_0 \left[2 l(l+1)-x_0^2 \ddot{f}(x_0)\right],
\label{stu_series}
\eeq
where we have used $\dot{f}(x_0)=0$ and
\be
\ddot{f}(x_0)=x_0^{-4} \left.\f{d^2 f}{dr^2}\right|_{r_0}=\si x_0^{-4}
\ee
with the second moments $\si=(d^2 f/dr^2 )|_{r_0}$ given by (\ref{sigma}) \footnote{For the black hole cases, the coefficients are obtained as
$s_{0} =2 \ka x_{+}^2,~
s_{1} = x_{+} [8 \ka-\f{1}{2}x_{+}^3 \ddot{f}(x_{+}) ],~
t_0=2 x_{+}^2 ( \ka-i \om ),~
t_1=-x_{+} [-12 \ka+ x_{+}^3 \ddot{f}(x_{+})+4i \om ],~
t_2=-2i \om+ ~{({\rm non-}\om~ {\rm  terms})}, ~
t_{n(>2)}=~{({\rm non-}\om~ {\rm  terms})},~
u_1 = l(l+1) x_{+}^2 +2\ka x_{+} + m^2, ~
u_2 =x_{+} [2 l(l+1)+6 \ka x_{+}^{-1}-x_{+}^2 \ddot{f}(x_{+})]$.
{Compared} with (\ref{stu_series}), the most important qualitative difference is the absence of $s_{-1}$ in the black hole case so that the horizon $x_+=1/r_+$ becomes a regular singular point. Other
coefficients look similar, with the role of the surface gravity $\ka$ at the black hole horizon replaced by $x_0 f(x_0)$ in our natural wormhole case:
This may be understood from
the direct relation
$x_0 f(x_0)=2 \ka|_{r_+ \ra r_0}=4 \pi {T_H}|_{r_+ \ra r_0}$ in (\ref{f:Eq}).}.

Now in order to consider the expansion of the solution around the throat $r_0$, we first set $\varphi=(x-x_0)^{\al}$ as the lowest order solutions. Then, at the leading {order} $n=-1$, one obtains the indicial equation,
\be
s_{-1} \al (\al-1)=-x_0^4 f(x_0) \al (\al-1)=0,
\label{indicial_eq}
\ee
which {gives} two solutions, $\al=0$ and $\al=1$. The first solution, $\al=0$,
corresponds to the ingoing mode $\Phi \sim e^{-i\om v}$ near the throat.
The second solution, $\al=1$, is also an ingoing mode near the throat but
{\it vanishing} at the throat as $\Phi \sim e^{-i\om v}(x-x_0)$. Since in
this paper we want to consider non-vanishing ingoing modes to study QNMs,
we take only the $\al=0$ case
\footnote{The $\al=1$ case corresponds to {\it normal modes} without
any wave flow, {\it i.e.,} energy loss
 at the throat. It is interesting
{that} our wormhole system allows  also this solution as well as QNMs with the ingoing mode solutions of $\al=0$. This is in contrast to the black hole system, where outgoing mode solutions are allowed instead, as well as the ingoing modes at the horizon.}. Then the desired solution can be expanded as
\be
\varphi(x)=\sum_{n=0}^{\infty}a_{n}(x-x_0)^n.
\label{varphi_expansion}
\ee
{Plugging} (\ref{varphi_expansion}) into (\ref{Rad_Eq_EF_x}) with the expansion (\ref{stu_expansion}), one obtains the recursion relation for $a_n~(n \neq 0,1)$ as {follows:}
\be
a_n=-\f{1}{P_n}\sum_{k=0}^{n-1}
\left[k(k-1)s_{n-k-1}+kt_{n-k-1}+u_{n-k-1}\right]a_k ,
\label{Recursion_rel}
\ee
where
\be
P_n=n(n-1)s_{-1}=-n(n-1) x_0^4 f(x_0).
\ee
{Generally we can get two parameter families of solutions in terms of $a_0$ and $a_1$ near $x=x_0$. }
As we have discussed already, $a_0$ term corresponds to {
a pure ingoing mode} at $x=x_0$ so that $a_0$ should be kept
{and in this paper we set $a_0=1$ for
{convenience}}. On the other hand, $a_1$ term corresponds to {a vanishing} ingoing mode at $x=x_0$ so that we {may} discard this family of solution, which actually satisfies {the} additional Neumann boundary condition for $\varphi$, {\it i.e.,} $\varphi '|_{x_0}=a_1 \equiv 0$. In this paper, we are only interested in this case for simplicity.

As $r \ra \infty$, (\ref{Rad_Eq_EF_x}) reduces to
\be
r^2\f{d^2 \varphi(r)}{dr^2}+ 2 r \f{d \varphi (r)}{dr}-\left(2-\f{3 m^2}{\La}\right) \varphi(r) \approx 0,
\ee
which leads to the asymptotic solution as
\be
\varphi(r) \approx C_1 r^{-\f{1}{2} (1+\sqrt{9-12 m^2/\La})}+ C_2 r^{-\f{1}{2}(1-\sqrt{9-12 m^2/\La})}
,
\label{sol_tort_infy}
\ee
corresponding to the solution in (\ref{sol_tort_infy_tilde}).
Since we are interested in the normalizable modes, we take
{$C_2=0$} and the
{desired} solution {is} $\varphi(r) \sim r^{-\f{1}{2}(1+\sqrt{9-12 m^2/\La})}$, which also satisfies the vanishing Dirichlet boundary condition $\varphi(r) \ra 0$ as $r\ra \infty$ ($x \ra 0$). This means that we impose {the boundary condition as an algebraic equation} at $x=0$,
\be
\varphi(0)=\sum_{n=0}^{\infty}a_{n}(-x_0)^n=0,
\label{D_BC}
\ee
which is satisfied {only} for some discrete values of $\om$
since $a_n$'s are functions of $\om$
{from} (\ref{Recursion_rel}). If the sum (\ref{D_BC}) is
convergent, one can truncate the summation at some large
{order} $n=N$ where the partial sum beyond $n=N$ does not
 change within the desired precision. Because this approach can easily
 be implemented numerically, particularly in {\it Mathematica}, the
 coefficients $a_n, s_n, t_n,u_n$ can be computed up to an arbitrary
 order $N$. In the next section, we present the numerical computation of
QNM frequencies $\om \equiv \om_R-i \om_I$ based on this method.

\section{Numerical Results and Their Interpretations}

In this section, we will show the results of numerical computation of QNMs as described in the previous section. {In this paper, we
consider only the ``lowest" QNMs, whose absolute magnitude,
$| \om|=\sqrt{\om_R^2+\om_I^2}$, is the smallest, unless stated otherwise.}
First of all, Fig. 6 and 7 show the lowest QNM
frequency $\om=\om_R-i \om_I$ as a function of $m^2$ for
{varying}
$M$ and $Q$. Here, we focus on the case of $\beta Q>1/2$, where the well-defined
GR limit of $\beta \ra \infty$ exists \cite{JKim:2016}. The result shows that
the perturbations are stable ($\om_I>0$) if $m^2$ is above certain threshold
values $m_*^2$ ($-0.16769, -0.16720, -0.16704$ for $M=10^{-5}, 0.1, 0.15$
with $Q=1$ (Fig. 6); $-0.16769, -0.16728, -0.16704$ for $Q=1, 0.9, 0.85$
with $M=10^{-5}$ (Fig. 7)). For a given value of $M$ or $Q$, {QNM frequencies $\om_I$ and
$\om_R$} {increase} as $m^2$ {increases} above $m_*^2$. Here, we note that the
critical mass $m_*^2$ is close to the BF bound of (\ref{BF_bound}),
$m^2_{\rm{BF}}=-3 \La/4=-0.15$ \cite{Brei:1982}.
On the other hand, for a given {value of} $m^2$, $\om_I$ and $\om_R$ increase as $M$ decreases or $Q$ increases, {corresponding} to  increasing throat radius $r_0$ from (\ref{M_r0}) (Fig. 2). Neglecting the small differences of $m_*^2$ for different $M$ and $Q$, we may approximately fit the numerical result of the $m^2$ dependence in Fig. 6 and 7 {to} the {\it analytic} functions
\beq
&&\om_I \approx b_1 (m^2-m_*^2)+ b_2 (m^2-m_*^2)^2+\cdots, \no\\
&&\om_R \approx \f{1}{2}+c_1 (m^2-m_*^2)+ c_2 (m^2-m_*^2)^2+\cdots,
\label{omega_near_critical}
\eeq
near the critical mass squared $m_*^2$, {with the
approprite coefficients, $b_n$ and $c_n (n=1,2, \cdots).$}

\begin{figure}
\includegraphics[width=6cm,keepaspectratio]{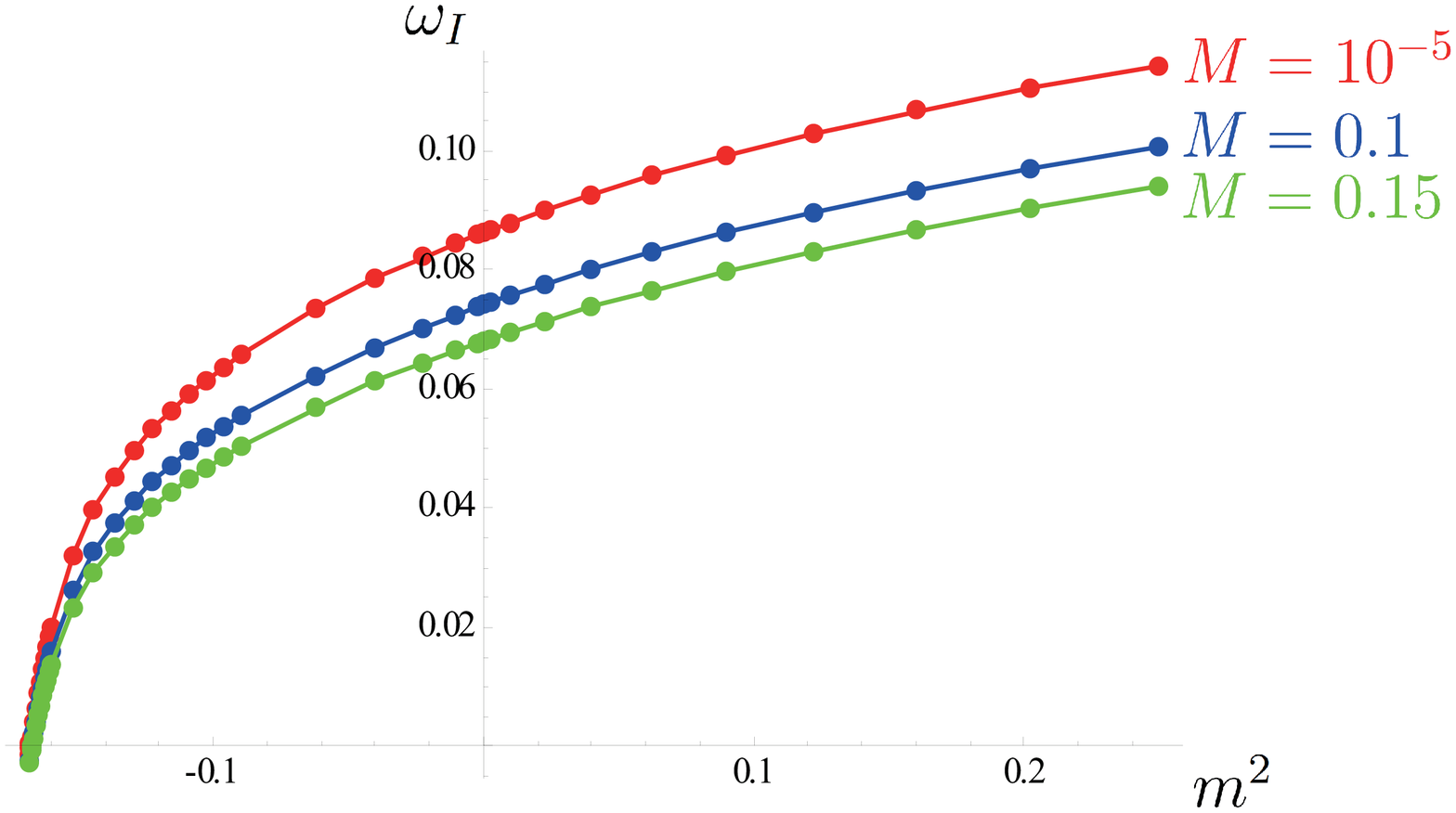}
\qquad
\includegraphics[width=6cm,keepaspectratio]{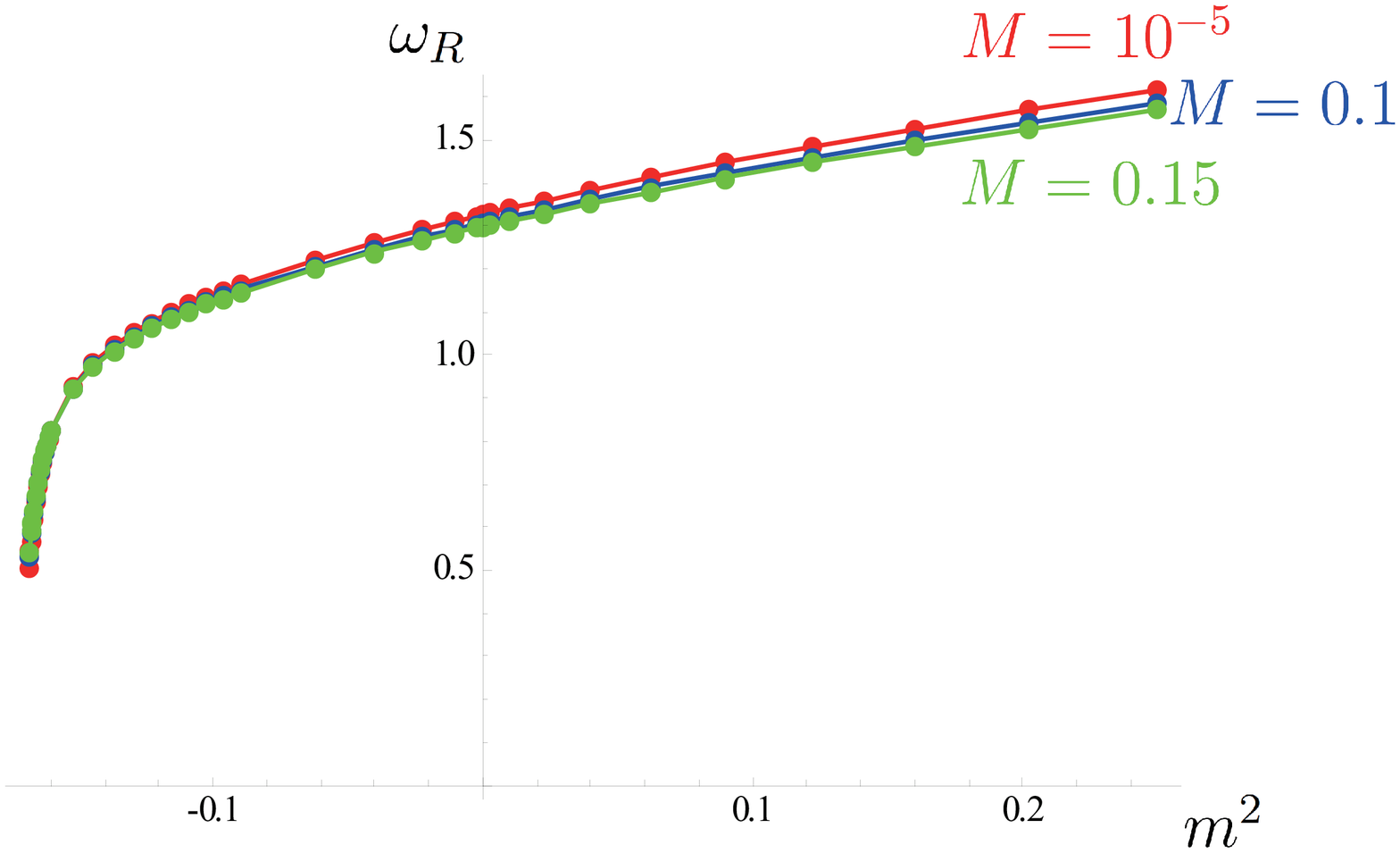}
\caption{The plots of $\om=\om_R-i \om_I$ vs. $m^2$ for {
varying}
$M$ with a fixed $\beta Q >1/2$ and $\La<0$. Here, we consider $l=2, \beta=1,Q=1,{\La=-0.2}$, and $M=10^{-5}, {0.1, 0.15}$ (top to bottom curves). The { result shows the instability ($\om_I<0$)} of massive perturbations below certain (threshold) values of mass squared, $m^2 <m_{*}^2$ with $m_{*}^2 \approx {-0.16769, -0.16720, -0.16704}$, respectively.}
\label{fig:QNM_varying_M_vs_m}
\end{figure}

\begin{figure}
\includegraphics[width=6cm,keepaspectratio]{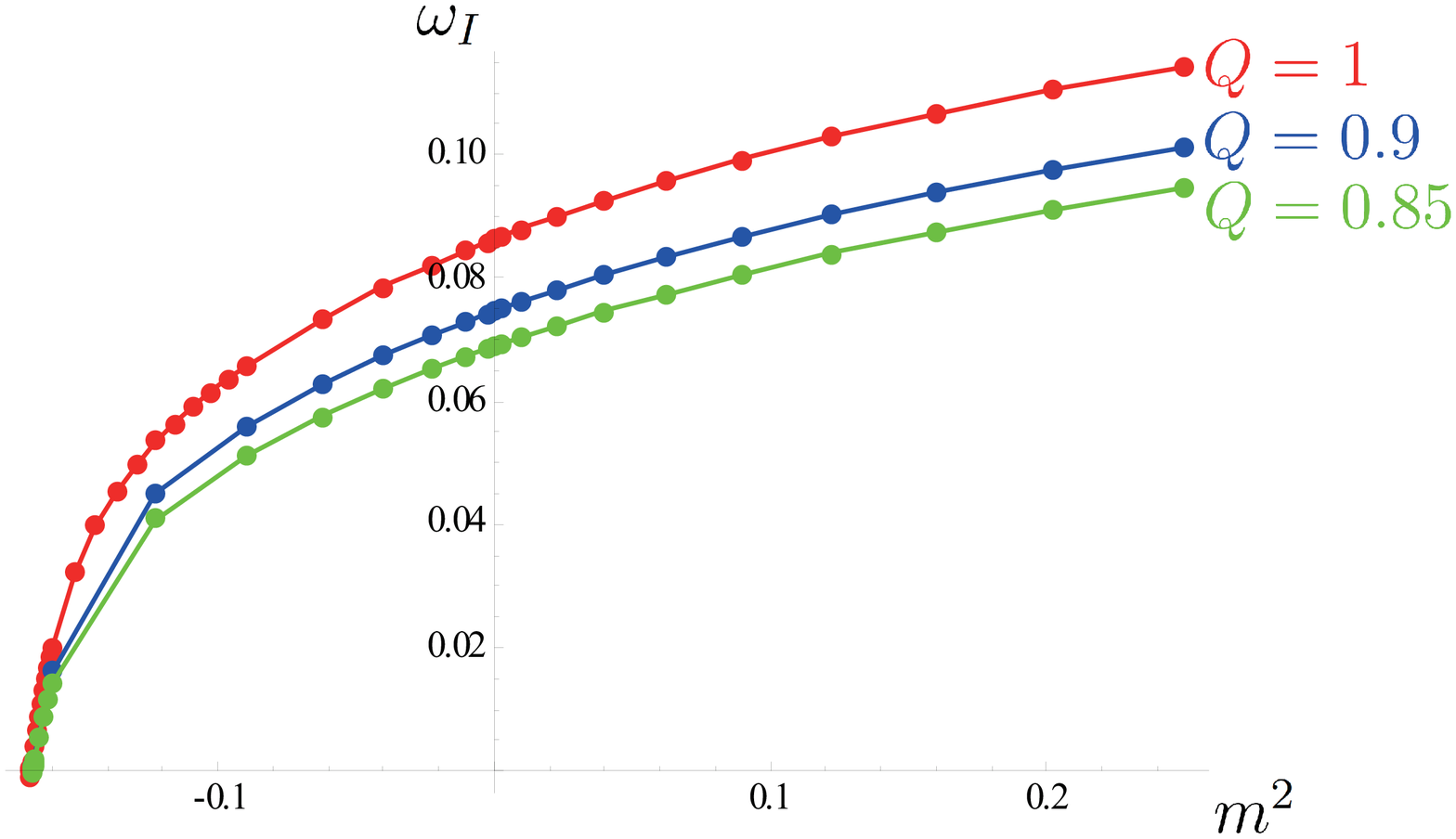}
\qquad
\includegraphics[width=6cm,keepaspectratio]{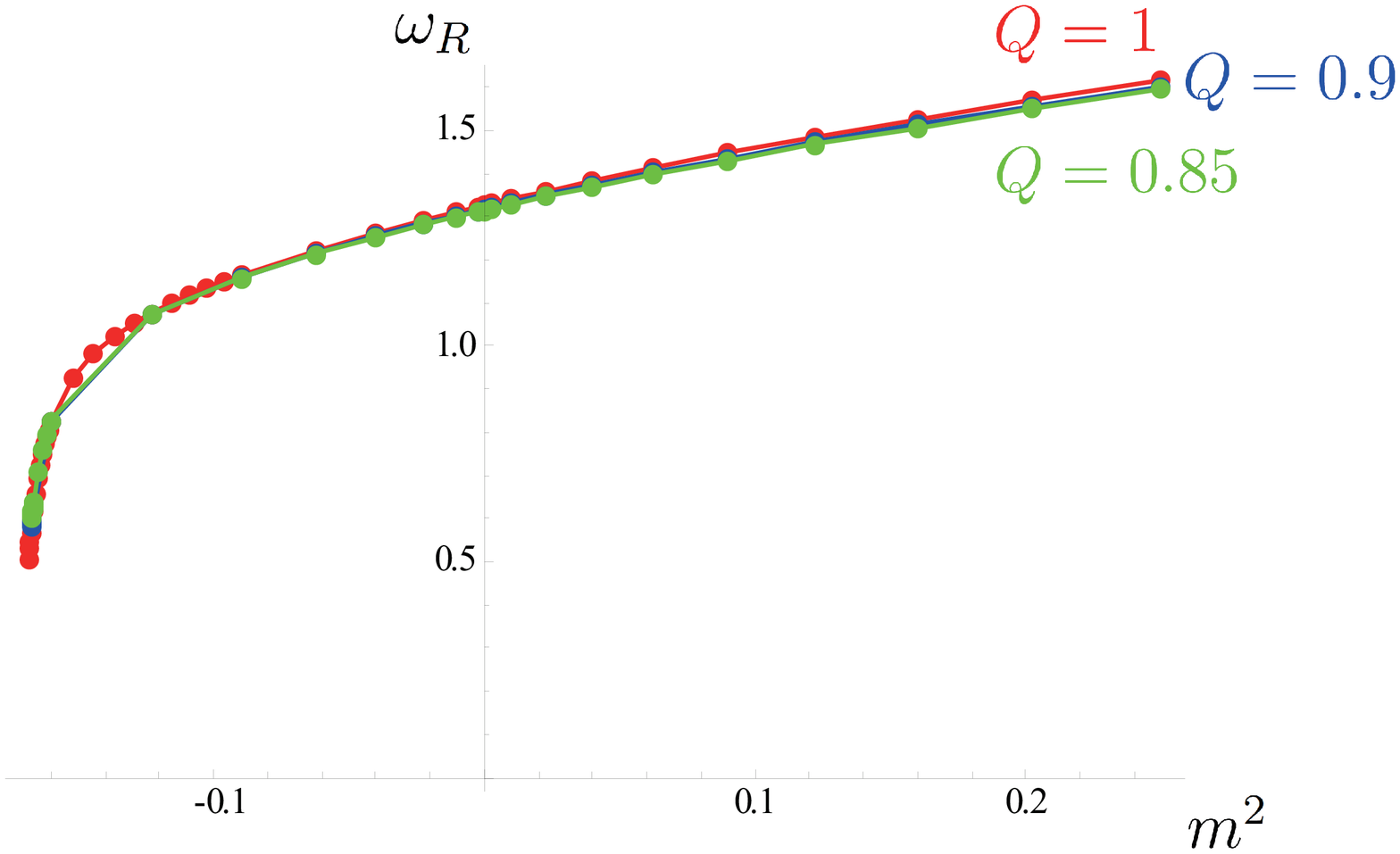}
\caption{The plots of $\om=\om_R-i \om_I$ vs. $m^2$ for {
varying}
$Q$ with a fixed $\beta Q >1/2$ and $\La<0$. Here, we consider $l=2,\beta=1,{\La=-0.2}, M=10^{-5}$, and $Q=1, 0.9, 0.85$ (top to bottom). The { result shows nearly the same on-set point of instability} as in Fig. 6, {$m^2_* \approx -0.16769, -0.16728, -0.16704$, respectively.}}
\label{fig:QNM_varying_Q_vs_m}
\end{figure}

Fig. 8 shows the QNM frequency as a function of $m^2$ for
{varying} negative cosmological constant $\La$ with fixed $M$ and $Q$. The result shows that the critical mass squared shifts as $m_*^2 \approx -0.16769, -0.11156,-0.08294$ for $\La=-0.2,-0.13,-0.1$, respectively. {This is consistent with the shifts of the corresponding $BF$ bounds, $m_{\rm{BF}}^2=3 \La/4=-0.15,-0.1,-0.075$. }

\begin{figure}
\includegraphics[width=6cm,keepaspectratio]{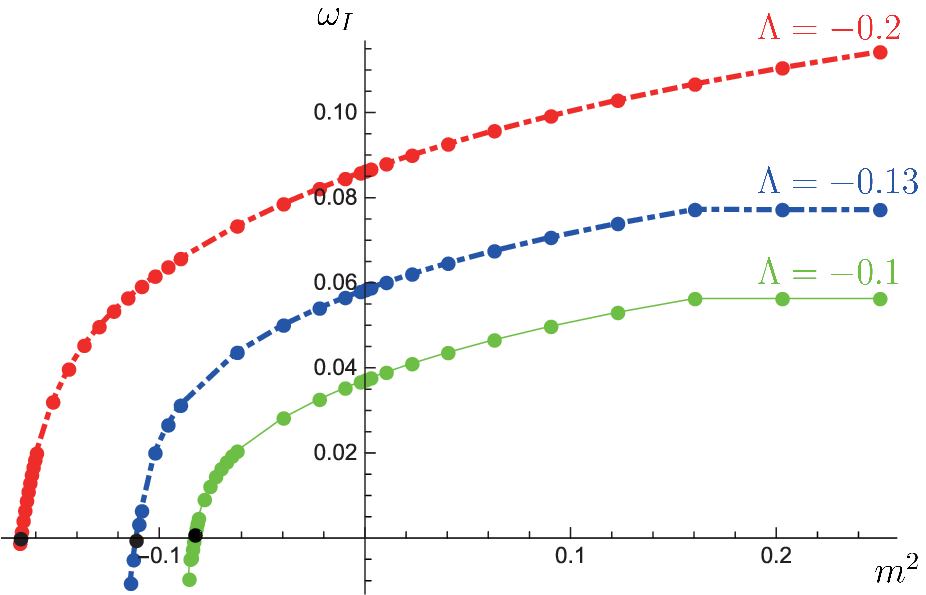}
\qquad
\includegraphics[width=6cm,keepaspectratio]{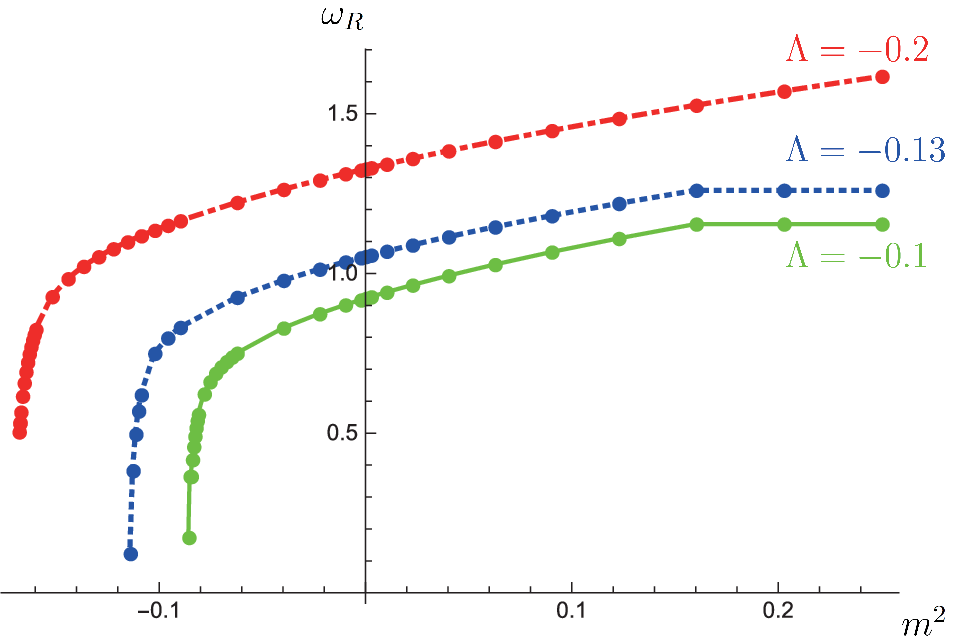}
\caption{The plots of $\om=\om_R-i \om_I$ vs. $m^2$ for {
varying}
$\La <0$ with a fixed $\beta Q >1/2$. Here, we consider $l=2,\beta=1,M=10^{-5},Q=1$, and $\La=-0.2,-0.13,-0.1$ (top to bottom). The {result shows the shifts of on-set points of instability} as $m_*^2 \approx -0.16769, -0.11156,-0.08294$ for $\La=-0.2,-0.13,-0.1$, respectively.}
\label{fig:QNM_varying_Lambda_vs_m}
\end{figure}

Fig. 9 shows the {lowest} QNM frequency
as a function of $m^2$ for
{varying} angular-momentum number $l$. The result shows that the critical mass shifts as
$m_*^2 \approx -0.16769, -0.16679,-0.15524$ for $l=2,1,0$, respectively.
{Especially for $l=0$, it shows a {\it level crossing} between the lowest {\it pure-imaginary} mode representing an over-damping (purple color) and the lowest {\it complex} mode (pink color) at
$m^2\approx -0.1089$ before the critical mass $m^2_*$ is being reached, so that the instability is governed by the {used-to-be higher} (pure) imaginary frequency modes \footnote{In this paper we have so far considered only the lowest QNMs or the small region around the level-crossing point. In order to discuss higher modes or the larger region around the crossing point, we need to increase the order $N$ since the numerical accuracy decreases as the mode number is increased generally. It would be interesting to check whether other unstable branches exist for higher modes but this is beyond the scope of this paper.}. On the other hand, there are discontinuities in $\om_I$ and $\om_R$ at $m^2 \approx -0.0784$ for the lowest {\it complex} frequency $\om=\om_R-i \om_I$ (green color) even before the level-crossing point \footnote{For the level crossings in black strings, where the effective {masses due to Kaluza-Klein reduction is} naturally introduced, see \cite{Kono:2008}. It seems this phenomena is universal in massive QNMs.}.}

\begin{figure}
\includegraphics[width=6cm,keepaspectratio]{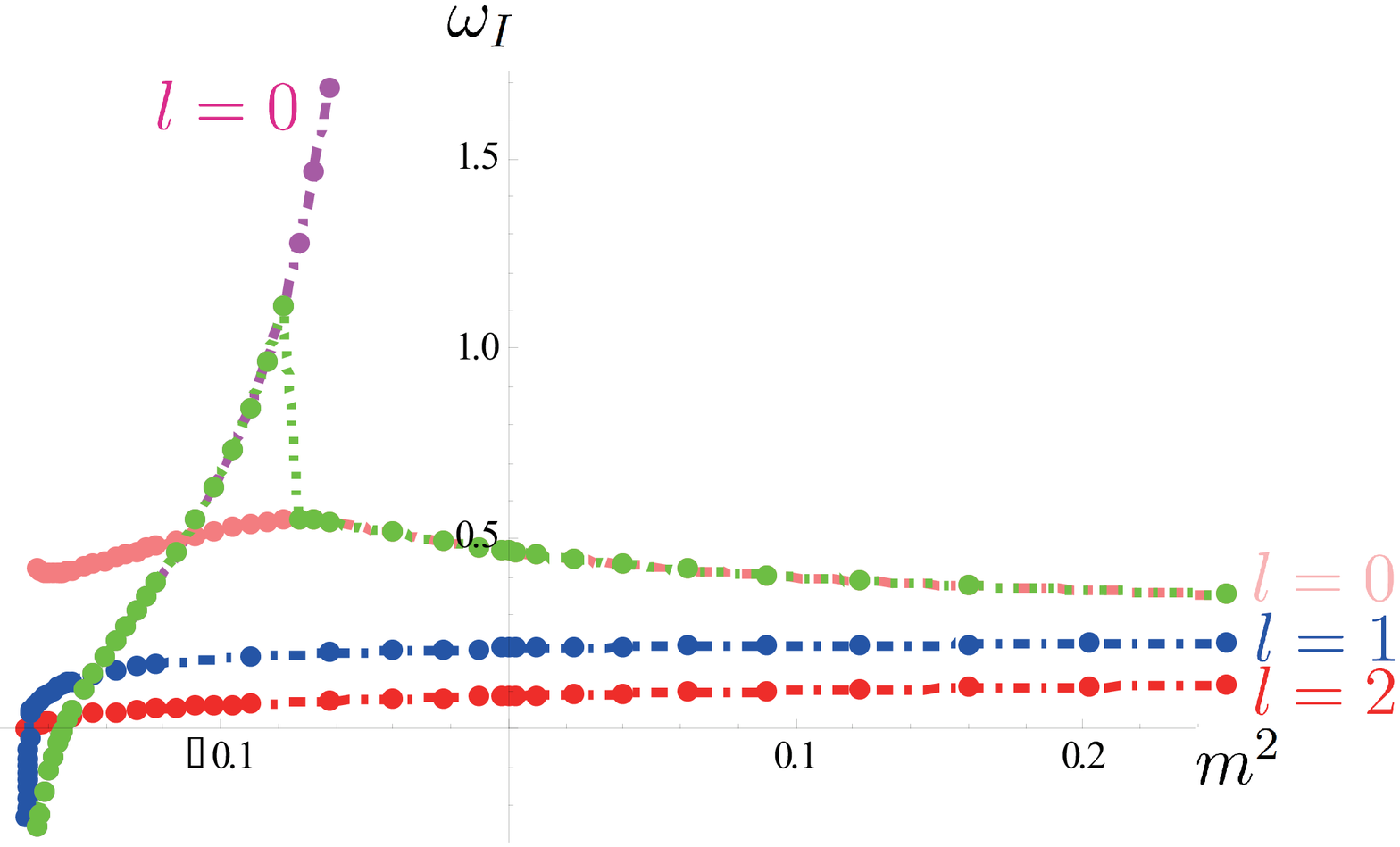}
\qquad
\includegraphics[width=6cm,keepaspectratio]{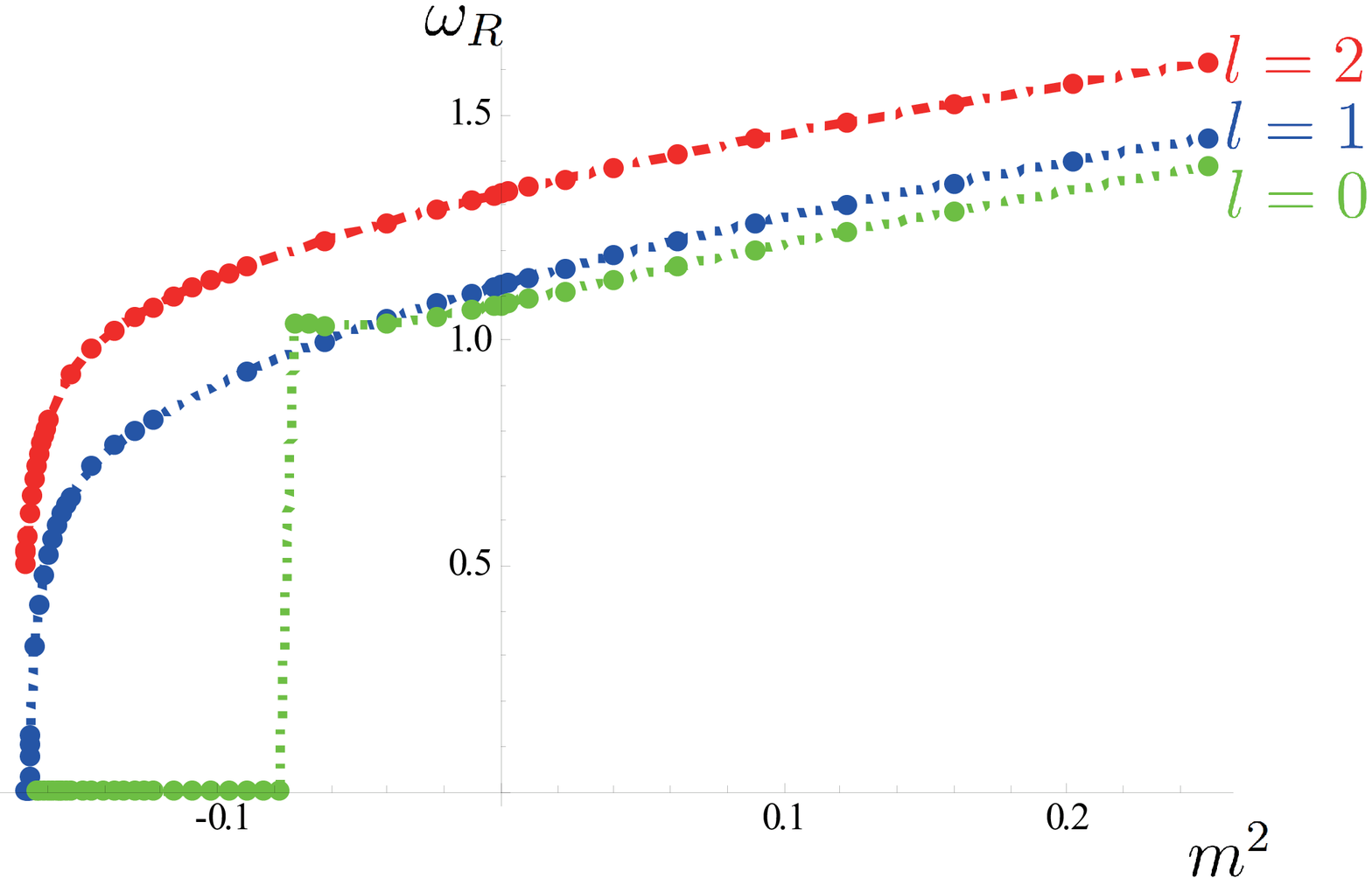}
\caption{The plots of $\om=\om_R-i \om_I$ vs. $m^2$ for {
varying}
$l$
(for left, $l=0$ (top two curves), $l=1$ (second from bottom), $l=2$
(bottom); for right, $l=0,1,2$ (bottom to top)) with a fixed $\beta Q >1/2$
and all other parameters. Here, we consider
$\beta=1,Q=1,M=10^{-5},\La=-0.2$.  The result {shows that} the critical mass shifts as
$m_*^2 \approx -0.16769, -0.16679,-0.15524$ for $l=2,1,0$, respectively.
{ When $l=0$, it shows a {\it level crossing} between
the lowest {\it pure-imaginary} mode
(purple color) and the lowest {\it complex} mode (pink color) at $m^2\approx -0.1089$
before the on-set of instability is being reached. Moreover, there are discontinuities in $\om_I$ and $\om_R$
for the lowest $\om=\om_R-i \om_I$ (green color) at $m^2\approx -0.0784
$ before the level-crossing point. }
 }
\label{fig:QNM_varying_l_vs_m_1}
\end{figure}

Fig. 10 shows the QNM frequency as a function of $l$ for
{varying} $m^2$. For small $l$ ($l \leq 5$), $\om_I$
decreases but $\om_R$ increases (except for the case $m^2<0$) as $l$ increases, whereas for large $l$, both $\om_I$ and $\om_R$ increase as $l$ increases. It is interesting {to note} that there is a bouncing point of $\om_I$, where $\om_I=0$ at $l=5$. Here we have considered only the case  $m^2 >m_*^2$, which is stable for small $l$.

\begin{figure}
\includegraphics[width=7cm,keepaspectratio]{{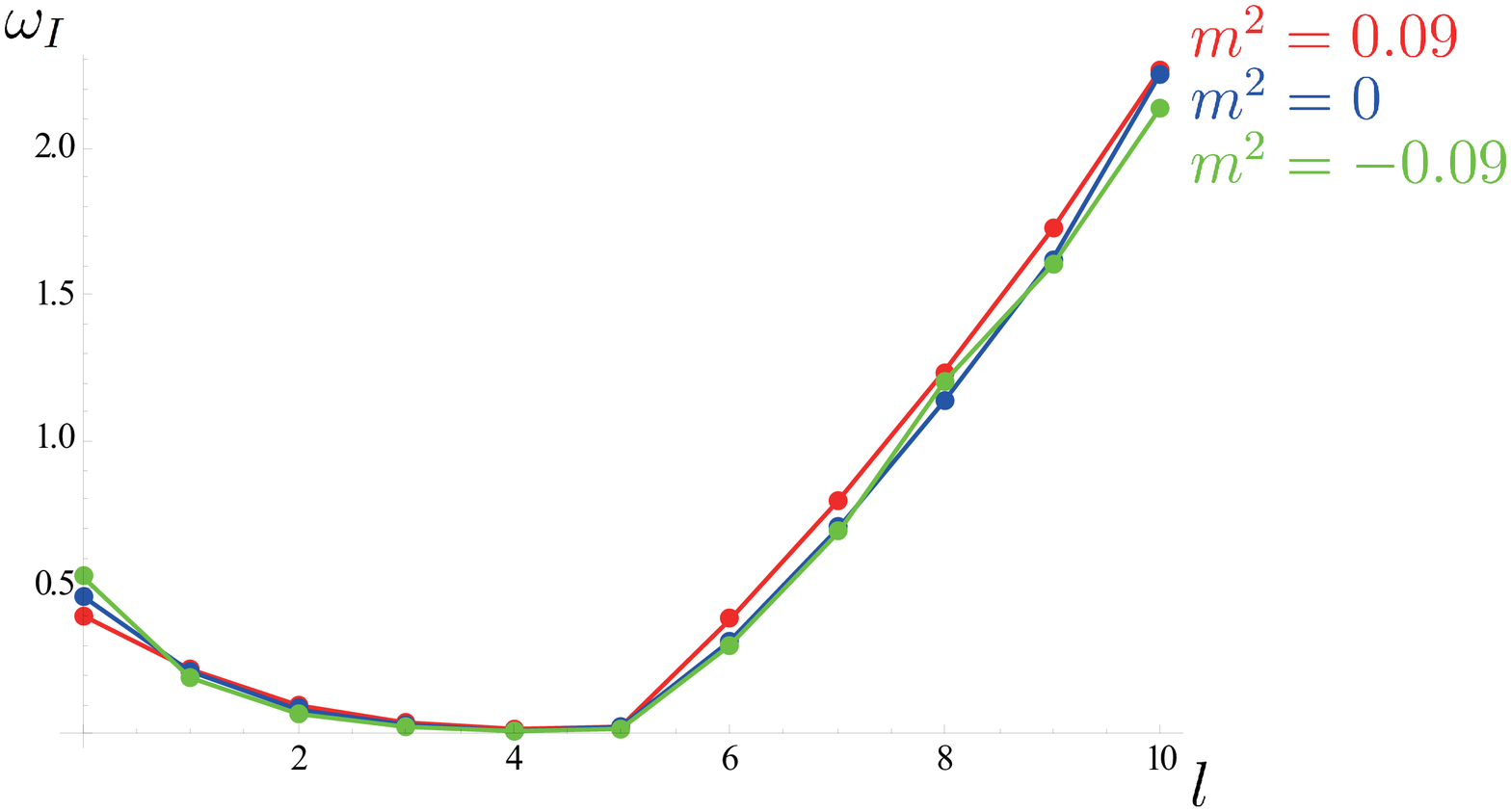}}
\qquad
\includegraphics[width=7cm,keepaspectratio]{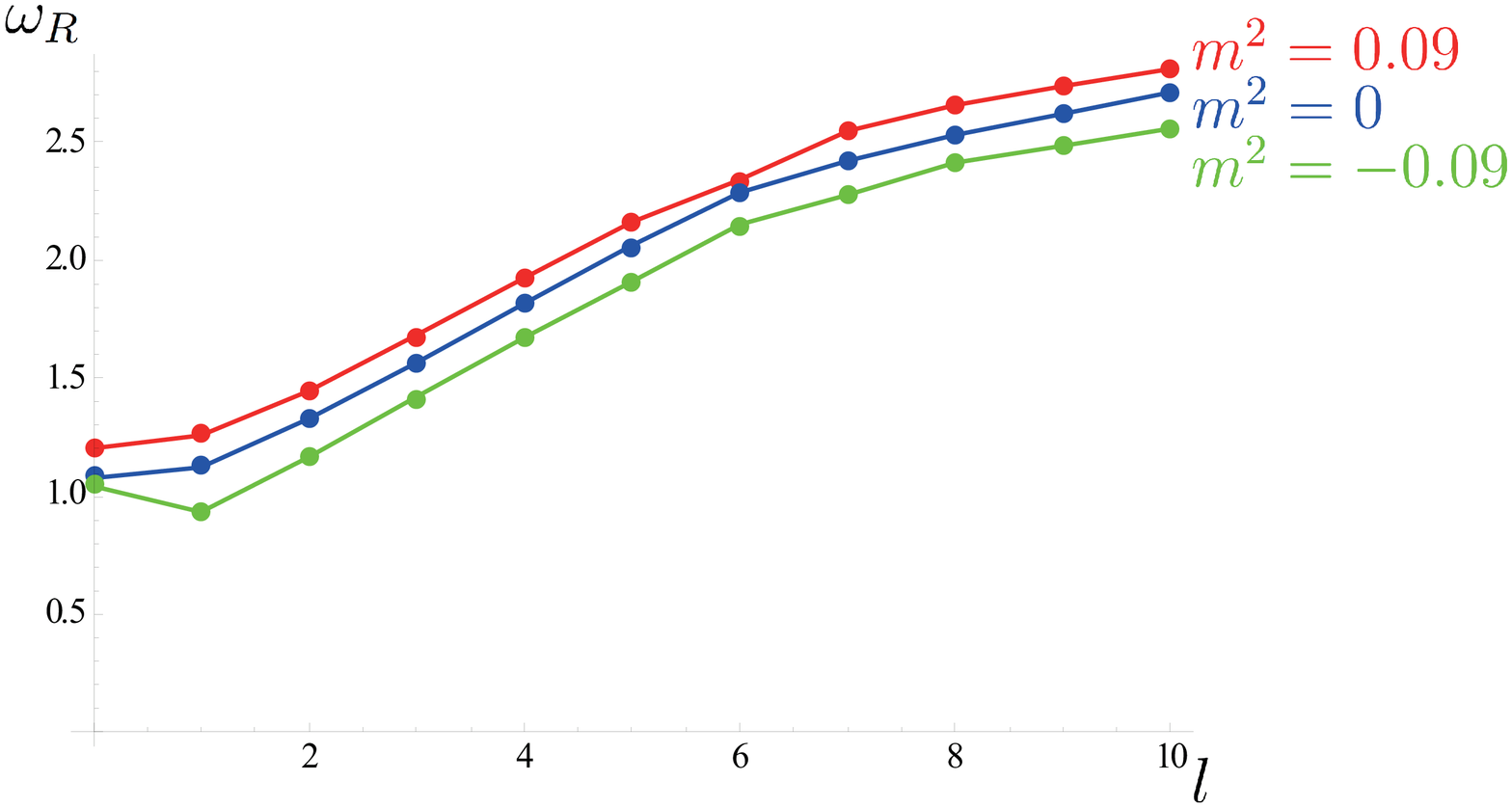}
\caption{The plots of $\om=\om_R-i \om_I$ vs. $l$ for {
varying}
$m^2$ ($m^2=0.09,0,-0.09$ (top to bottom)) with a fixed $\beta Q >1/2$ and all other parameters. Here, we consider $\beta=1,Q=1,M=10^{-5},\La=-0.2$. For small $l~ (\leq 5)$, $\om_I$ decreases as $l$ increases but $\om_R$ increases (except for the case $m^2<0$), whereas for large $l$ both $\om_R$ and $\om_I$ increase as $l$ increases. There exists a bouncing point of $\om_I$, where $\om_I=0$ at $l=5$. Here we consider only the case {$m^2>m_*^2$, which is stable for small $l$.}}
\label{fig:QNM_varying_l_vs_m_2}
\end{figure}

Fig. 11 shows the QNM frequency as a function of $r_0$ for
{varying} $m^2$ in Fig. 6. {Though the result
is preliminary since there are only limited data points, {\it i.e.}, for
a given value of $m$,
only three points of $r_0$ which correspond to the three values of $M$ in
Fig. 6, the result indicates interestingly the linear dependence of QNM
frequency to the throat radius.
 Explicitly, the curves can be fitted to
\beq
\om_I \approx d_0+d_1 r_0,~~
\om_R \approx e_0+e_1 r_0 \label{omega_linear_fit}
\eeq
with $d_0 = -0.15050, -0.14728, -0.13157,~ d_1 = 0.12718, 0.11894, 0.10046, ~ e_0 = 0.98624, 0.96175, 0.92391,~ e_1 = 0.23483, 0.18614, 0.12275$ for $m^2 \approx 0.09,0,-0.09$, respectively.
This is similar to the case for large black holes where QNM frequency depends
linearly on the horizon radius ($\om_I, \om_R \propto r_+$)
\cite{Horo:1999,Wang:2000}
\footnote{Since the quantity ${f(r_0)}/{r_0}$, which is given by
$
-\La r_0 +{1}/{r_0}-{Q^2}/{r_0^3}+{Q^4}/{4 \beta^2 r_0^7}+
{\cal O}(r_0^{-11}),
$
corresponds to {the} surface gravity for black hole case,
$2\ka=4 \pi T_H=(df(r)/dr)|_{r_+}$, as noted in {footnote} No. 7,
the wormhole's QNM frequencies (\ref{omega_linear_fit}) may also be fitted {to} $\om_I \approx d_0-d_1 \La^{-1} f(r_0)/r_0,~ \om_R \approx e_0-e_1 \La^{-1} f(r_0)/r_0$ for the leading order of large $r_0$.}.

The asymptotic linear dependence can also be understood as the result of the scaling behavior \footnote{The scaling for the (abbreviated) wormhole mass $M$, which differs from that of scalar {field mass} $m$, is due to the omitted Newton's constant $G${: $G$} has the dimension $[M]^{-1} [L]^3 [T]^{-2}$ and hence transforms as $G\rightarrow \alpha^2 G$ so that the ADM mass{, $M_{ADM}\sim GM$,} transforms as $M_{ADM}\rightarrow\alpha M_{ADM}$ as in (\ref{scaling_symm}).} \cite{Horo:1999,Greg:1993,Yin:2010},
\beq
&&r\rightarrow \alpha r,~t\rightarrow \alpha t,~\omega\rightarrow\omega/\alpha,~\beta\rightarrow\beta/\alpha, \no \\
&&m\rightarrow m/\alpha,~M\rightarrow\alpha M,~Q\rightarrow\alpha Q,~\Lambda\rightarrow\Lambda/\alpha^2,
\label{scaling_symm}
\eeq
by which the perturbation equation (\ref{Rad_Eq_EF_x}) is unchanged. This means that the QNM frequency $\om=\om_R-i\om_I $, which is a function of $r_0, \La, Q, \beta$, {and $m$, should} have the following form
\beq
\om \sim 
\La r_0 + 
\f{1}{r_0}
+ \beta
+
\f{Q^2}{r_0^3}+
\f{Q^2}{\beta^2 r_0^5}
+
r_0 m^2+\f{r_0}{\La} m^4
+\cdots
\eeq
in order to have the scaling $\omega\rightarrow\omega/\alpha$. 
For large $r_0$, the dominant terms are given by
\beq
\om \approx \bar{\zeta} \La r_0 +  \bar{\eta} \beta+
 \left(\bar{\delta} m^2+\bar{\gamma}\La^{-1}m^4 +\cdots \right) r_0,
\eeq
where $\bar{\zeta}, \bar{\eta}, \bar{\delta}$, and $\bar{\gamma}$ are
scale invariant {coefficients}, which {agrees with} the behavior of
(\ref{omega_linear_fit})
\footnote{Expanding near the critical mass squared $m^2_*$ and
comparing with (\ref{omega_near_critical}), one can
obtain the relations $\bar{\zeta} \La r_0 +  \bar{\eta} \beta \approx 1/2-r_0 {\cal M}(m^2_*)
,~  c_1-i b_1 \approx {\cal M}'(m^2_*), ~  c_2-i b_2 \approx {\cal M}''(m^2_*)/2$,
where ${\cal M}(m^2)\equiv\left(\bar{\delta} m^2
+\bar{\gamma}\La^{-1}m^4 +\cdots \right)$ .}.

\begin{figure}
\includegraphics[width=6cm,keepaspectratio]{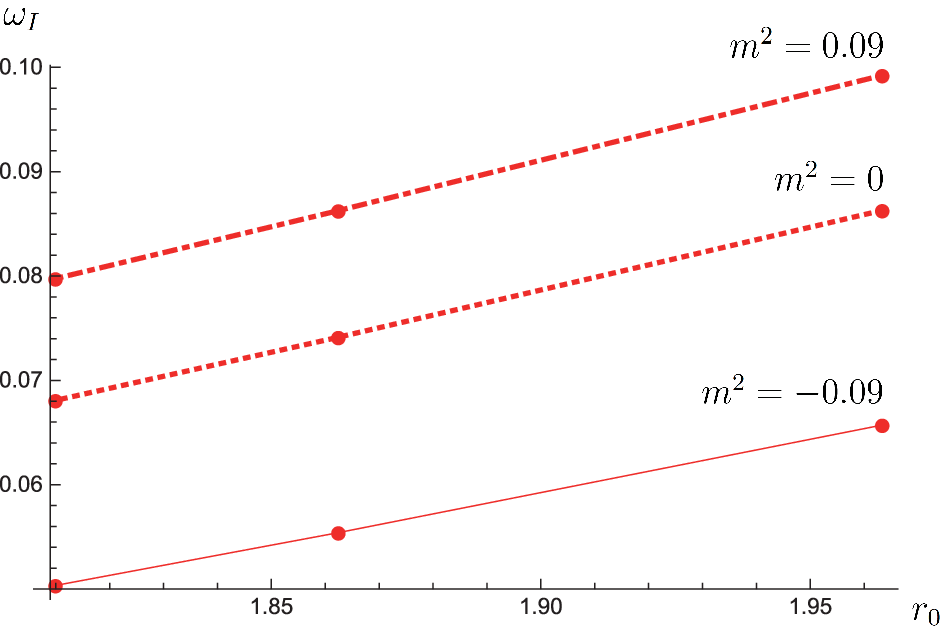}
\qquad
\includegraphics[width=6cm,keepaspectratio]{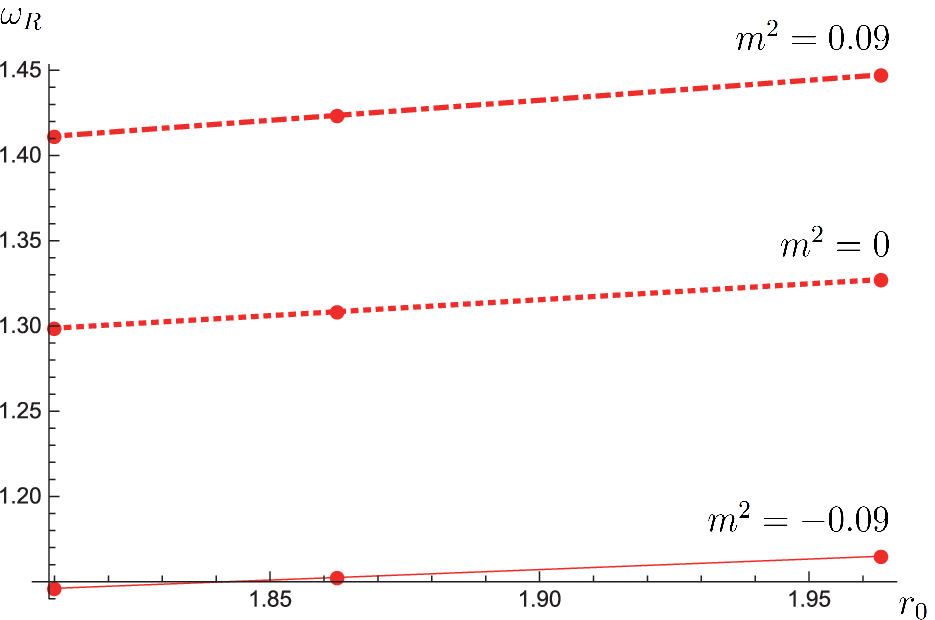}
\caption{The plots of $\om=\om_R-i \om_I$ vs. $r_0$ for
{
varying $m^2$ ($0.09,0,-0.09$ (top to bottom))} with a fixed $\beta Q >1/2$ and
all other parameters. Here, we consider three values of $r_0$ which
correspond to the three curves with $M=10^{-5}, 0.1, 0.15 \times 10^{-1}$
in Fig. 6.
The {result shows} the linear dependence of $\om_I, \om_R$ on
$r_0$.}
\label{fig:QNM_varying_m_vs_r_0}
\end{figure}

So far, we have studied the case $\beta Q>1/2$, which has a
well-defined GR limit at $\beta \ra \infty$. Now, we show in Fig. 12
the case $\beta Q \leq 1/2$, which do not have a GR limit. The result
shows that {there is no oscillatory part ($\om_R=0$) and moreover}
the critical mass squared $m_*^2<0$ does not occur in this case. Rather,
it shows another critical mass squared $m_c^2\approx 1.0326$ for
$\beta Q < 1/2$ so that the perturbations would be completely ``frozen",
{\it i.e.,} $\om=\om_R-i \om_I=0$, for $m^2>m_c^2$. Even though
this result could be preliminary too since we may not neglect the back
reaction of the wormhole geometry for the heavy-mass perturbations \footnote{For the case of $\beta Q>1/2$, the numerical accuracy decreases
as one increases the mass $m$ beyond the {plots} shown in
Fig. 12 and we did not include those cases in this paper.}
\cite{Kono:2004}, it seems to agree with the so-called
``{\it quasi-resonance modes (QRMs)}" with $\om_I=0$ in massive QNMs, though
{the oscillatory parts are different, {\it i.e.,} $\om_R=0$ in our wormhole case but $\om_R \neq 0$ in QRMs}
\cite{Ohas:2004}.

\begin{figure}
\includegraphics[width=8cm,keepaspectratio]{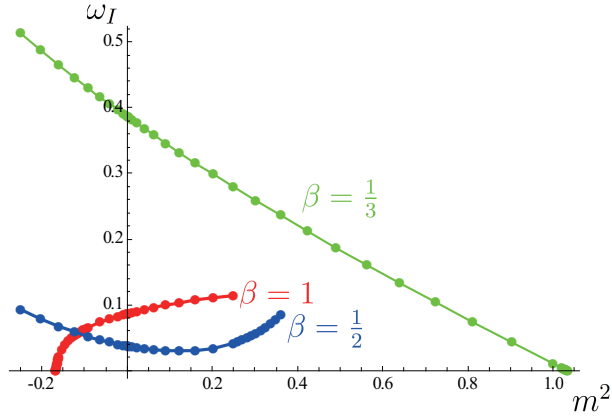}
\caption{The plots of $\om=\om_R-i \om_I$ vs. $m^2$ for
{
varying}
$\beta$ with a fixed $Q$ and all other parameters.
Here, we consider $l=2,Q=1,M=10^{-5},\La=-0.2$. The on-set of instability for $m^2<0$  occurs only for $\beta Q>1/2$ which
{was considered} in Fig. 6 and 7. For $\beta Q<1/2$, it
{shows
another on-set of instability}
at $m^2\approx 1.03326$. Moreover, for $\beta Q \leq 1/2$,
{the result shows only the pure-imaginary modes ($\om_R=0$).}
For $\beta Q > 1/2$, $\om_R$ {was} shown in Fig. 6.}
\label{fig:QNM_varying_beta_vs_m}
\end{figure}

\begin{figure}
\includegraphics[width=6cm,keepaspectratio]{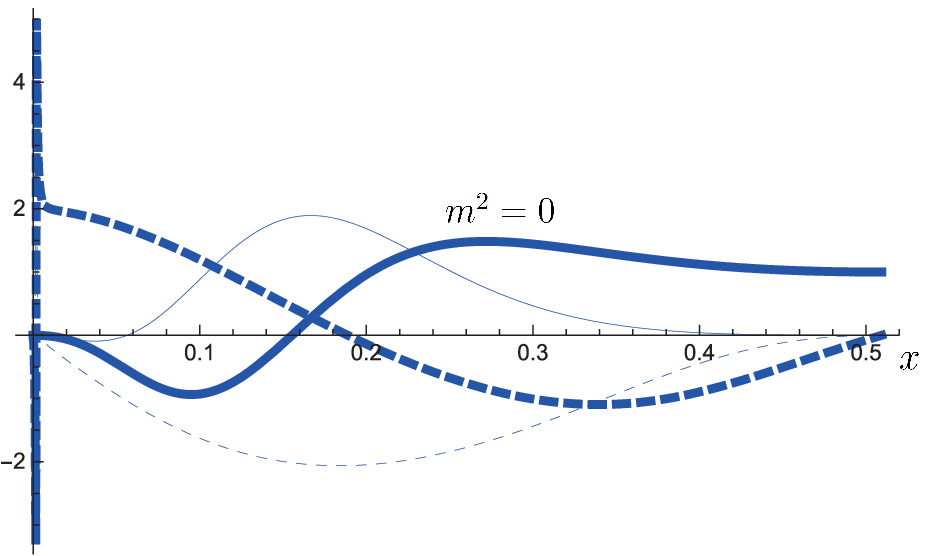}
\includegraphics[width=6cm,keepaspectratio]{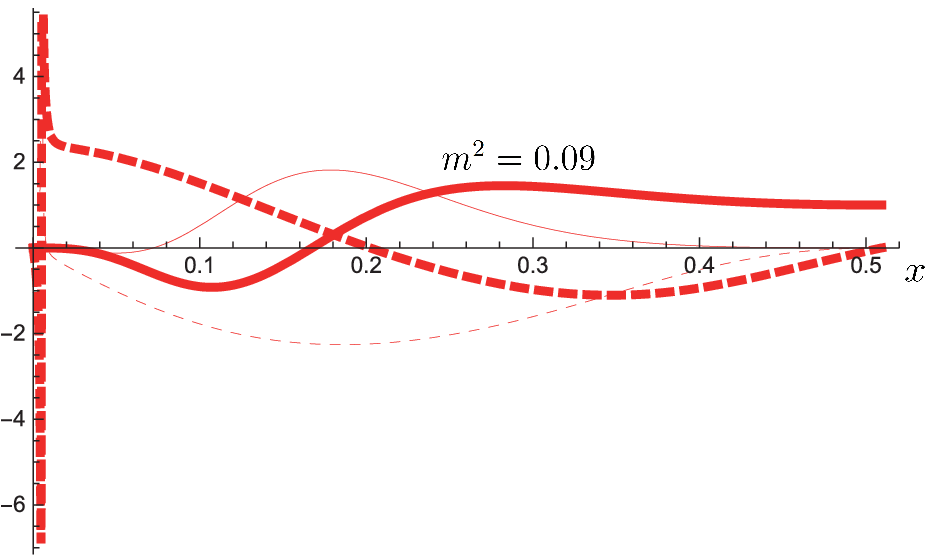}
\includegraphics[width=6cm,keepaspectratio]{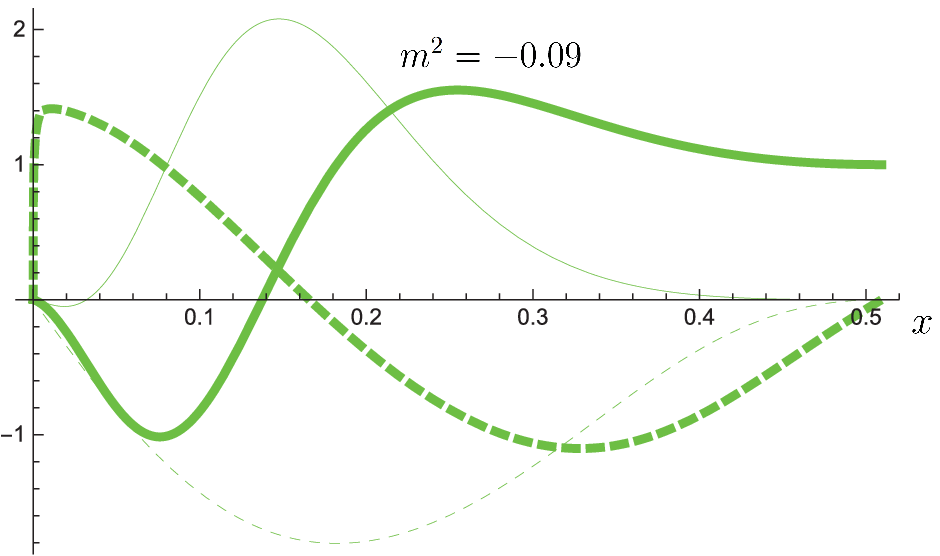}
\includegraphics[width=6cm,keepaspectratio]{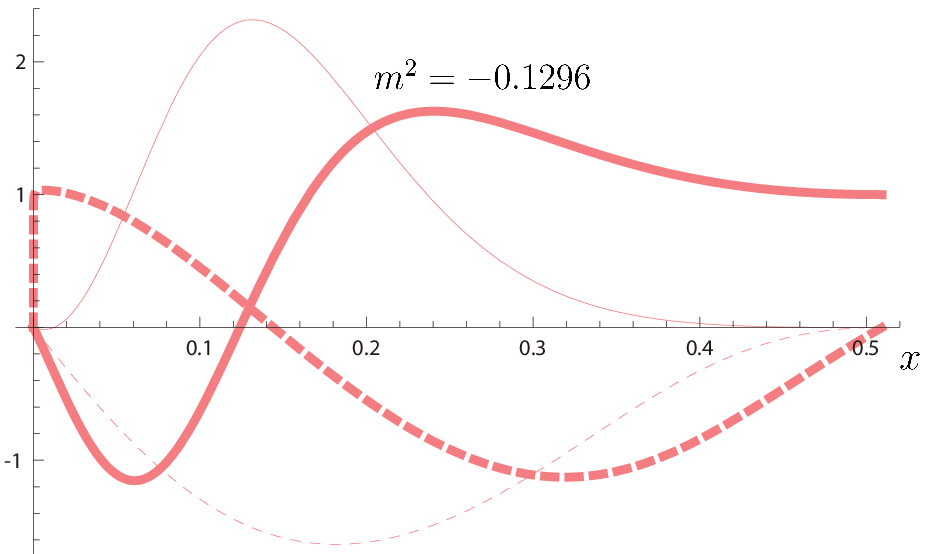}
\includegraphics[width=6cm,keepaspectratio]{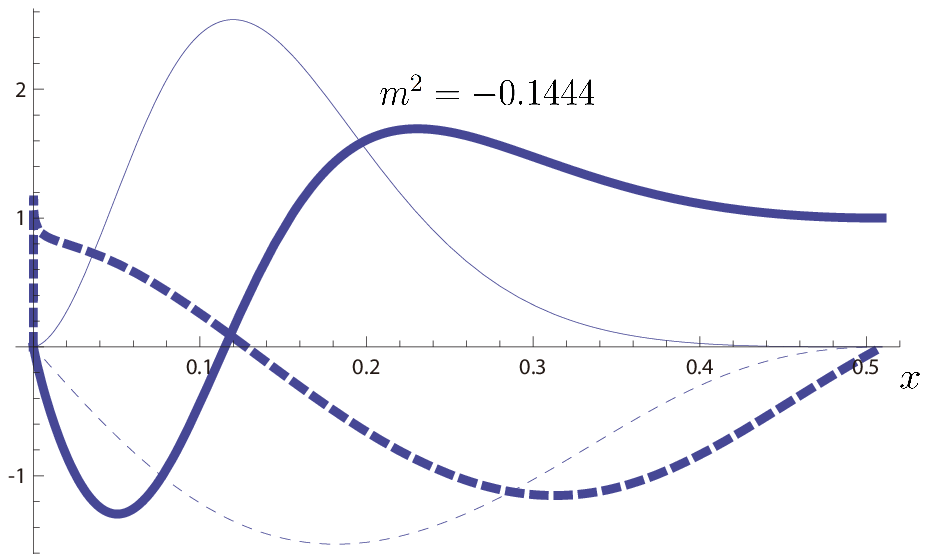}
\caption{The plots of the numerically {reconstructed (truncated)} wave function
$\varphi_N $ vs. $x$ for {
varying}
$m^2$ ($0, 0.09, -0.09,-0.1296,-0.1444$ (blue, red, green, pink, purple))
{and} $M=10^{-5}$ in Fig. 6.
The thick and thin solid lines denote $\mbox{Re}(\varphi_N)$ and $\mbox{Im}(\varphi_N)$, respectively and one can confirm that these satisfy the
vanishing Dirichlet boundary condition
{($\varphi(x) \ra 0$) as $x \ra 0$}
 and the vanishing Neumann boundary condition {
 ($d\varphi/dx \ra 0)$ as $x \ra x_0\approx 0.50939$}.
 The dotted lines denote their asymptotic exponents, $x \ln \mbox{Re}(d\varphi_N /dx),~
x \ln \mbox{Im}(d\varphi_N (x)/dx)$ and these {correctly} approach the exponent of our asymptotic solution $\varphi(x) \sim x^{\f{1}{2}(1+\sqrt{9-12 m^2/\La})}$ in (\ref{sol_tort_infy}), which are computed as $2,2.39737,1.44868,1.05317,0.78983$ for $m^2=0,0.09,-0.09,-0.1296,-0.1444$, respectively.}
\label{fig:QNM_varying_beta_vs_m}
\end{figure}

Before finishing this section, we end up with some remarks about the consistency of
our numerical results. First, Fig. 13 shows the {truncated} wave function $\varphi_N (x)$,
reconstructed from the numerically obtained $a_n$'s in (\ref{varphi_expansion}) up
to the order $N=300$. These show the vanishing Dirichlet boundary condition
($\varphi(x) \ra 0)$ as $x \ra 0$ and the
vanishing Neumann boundary condition ($d\varphi/dx \ra 0$) as
$x \ra x_0$ from our choice of $\alpha=1$ in the
indicial equation (\ref{indicial_eq}). 
Moreover, the asymptotic behavior of our desired wave function $\varphi(x) \sim x^{\f{1}{2}(1+\sqrt{9-12 m^2/\La})}$ in (\ref{sol_tort_infy}), {whose exponent can be captured by} $x \ln (d \varphi/dx) \ra \f{1}{2}(1+\sqrt{9-12 m^2/\La})$ as $x \ra 0$, {which are computed as $2,2.39737,1.44868,1.05317,0.78983$ for $m^2=0,0.09,-0.09,-0.1296,-0.1444$}, is well confirmed by the numerically reconstructed wave function \footnote{For $3 \Lambda/4 (\approx-0.15) \leq m^2 \leq 2 \Lambda/3 (\approx-0.13333)$,
the vanishing Dirichlet boundary condition {($\varphi(0)=0$)}
 may not uniquely {determine}
the desired solution with $C_2=0$ 
in (\ref{sol_tort_infy}), in contrast to the normalizability condition in Sec. III. But for a truncated summation, $\varphi_N (x)=\sum_{n=0}^{N}a_{n}(-x_0)^n$, it seems that only the more-rapidly decaying solution of $C_1$ part may be obtained from that boundary condition.}.

\begin{figure}
\includegraphics[width=6cm,keepaspectratio]{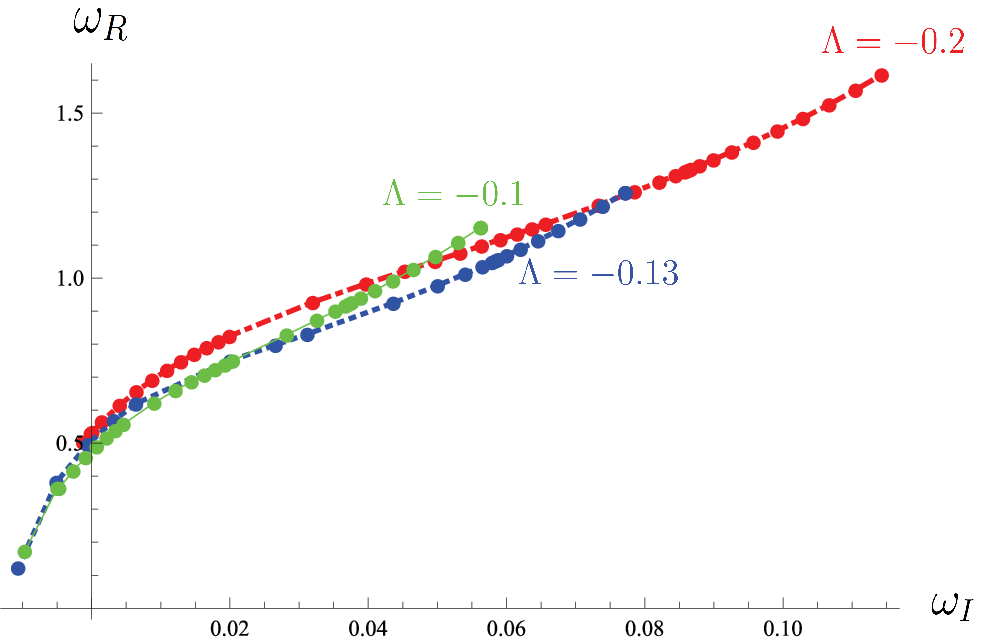}
\qquad
\includegraphics[width=6cm,keepaspectratio]{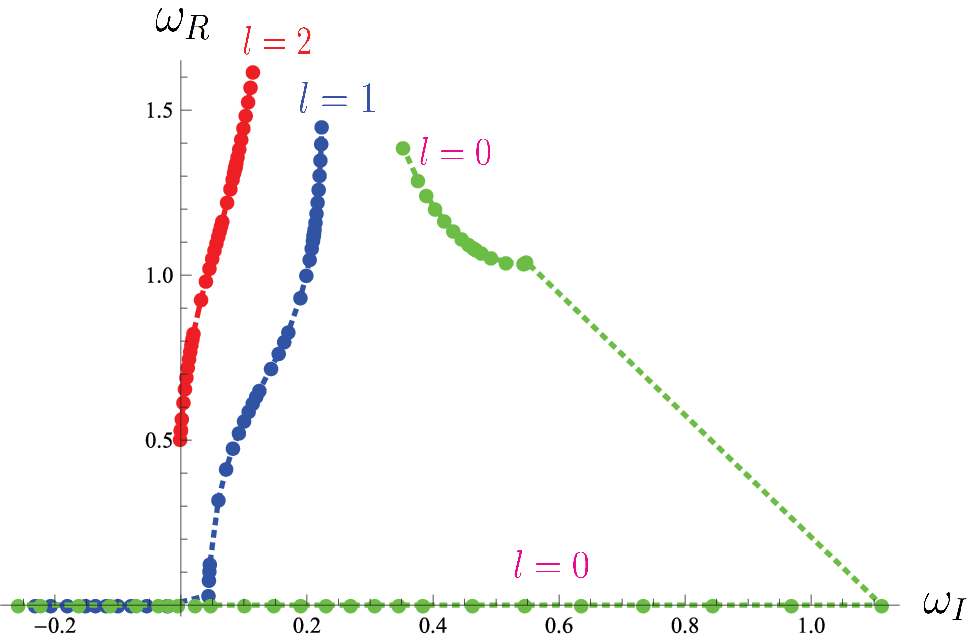}
\caption{{The plots of $\om_R$ vs. $\om_I$ for the results in Fig. 8 with
varying
$\La <0$ $(l=2)$ (left), and in Fig. 9 with
varying $l$ $(\La=-0.2)$ (right). The left shows a smooth non-vanishing function $\om_R=\om_R (\om_I)$ across the instability point for $\La=-0.13$ (blue) and $\La=-0.1$ (green). The right shows the vanishing $\om_R$ for the unstable region ($\om_I<0$) with non-smooth transitions before the instability point for $l=1$ (blue) and $l=0$ (green). The red curve is the {case} $\La=-0.2$ ($l=2$) in Fig. 8 and Fig. 9 shows also the smooth $\om_R=\om_R (\om_I)$ across the instability point though it is not quite clear in the plot.}}
\label{fig:QNM_varying_Lambda_vs_m}
\end{figure}

{Second, for the unstable modes ($\om_I<0$) beyond the critical mass $m^2<m^2_*$, Fig. 8 and 9 show that there exits the oscillation mode
with the non-vanishing
$\om_R$ which is
continuously changing across the on-set point of the instability
($\La=-0.13, -0.1$ case $(l=2)$ in Fig. 8 and $l=1$ case ($\La=-0.2$) in Fig. 9).
But
for some cases $\om_R$ vanishes with a sudden discontinuity before the on-set point of instability is being reached ($l=0$ case ($\La=-0.2$) in Fig. 9). This seems to contradict the argument in the literature which claims that ``
$\om_R=0$
for the unstable mode, $\om_I<0$ ", though there is no restriction on $\om_R$ for the stable mode, $\om_I>0$ \cite{Kono:2008,Moon:2011}. In order to clarify this issue, we consider the integral,
\beq
I=\int^{\de}_0 \left[ \widetilde{\varphi}^*
\f{d^2 \widetilde{\varphi}}{dr^2_*}+(\om^2-\widetilde{V}) |\widetilde{\varphi}|^2 \right] dr_* {=0}
\eeq
from the throat ($r_*=0$) {to} spatial infinity ($r_*=\de$), after multiplying {the complex conjugated function}
$\widetilde{\varphi}^*$ by (\ref{Rad_Eq_tort}). The
{partial integration
of the first}
gives
\beq
I=\left.\widetilde{\varphi}^*
\f{d \widetilde{\varphi}}{dr_*}\right|^{\de}_0+\int^{\de}_0 \left[
-\left|\f{d \widetilde{\varphi}}{dr_*}\right|^2+(\om^2-\widetilde{V}) |\widetilde{\varphi}|^2 \right] dr_*=0.
\eeq
Considering the desired wave function $\widetilde{\varphi}(x) \sim \widetilde{r}_*^{\f{1}{2}(1+\sqrt{1-4 \la})}$ with $B_2=0$ in (\ref{sol_tort_infy_tilde}), the boundary term at infinity $r_*=\de >0$ (or $\widetilde{r}_*=0$) vanishes \footnote{The boundary term vanishes as $\widetilde{r}_*^{\sqrt{1-4 \la}}$ for the desired solution. But it diverges as $\widetilde{r}_*^{-\sqrt{1-4 \la}}$ for the other solution $\widetilde{\varphi}(x) \sim \widetilde{r}_*^{\f{1}{2}(1-\sqrt{1-4 \la})}$ with $B_1=0$ in (\ref{sol_tort_infy_tilde}), which means the infinite amount of flux $\sim \widetilde{\varphi}^*
d \widetilde{\varphi}/dr_*$ at infinity. This can be considered as an alternative criterion for the desired solution \cite{Brei:1982,Birm:2001}.  } and one finds the imaginary part of the integral
\beq
{\rm Im} (I)=-{\rm Im}\left(\left.\widetilde{\varphi}^*
\f{d \widetilde{\varphi}}{dr_*}\right|_{r_*=0}\right)-2 \om_R \om_I \int^{\de}_0
 |\widetilde{\varphi}|^2 dr_*=0.
\eeq

When considering only the ingoing solution $\widetilde{\varphi}\sim e^{-ik r_*}$ with $A_2=0$ in (\ref{Phi_near_throat}), we now have
\beq
{\rm Im} (I)=k_R \left|\widetilde{\varphi}(r_*=0)
\right|^2-2 \om_R \om_I \int^{\de}_0 |\widetilde{\varphi}|^2 dr_*=0,
\eeq
from $({d \widetilde{\varphi}}/{dr_*})|_{r_*=0}=-ik \widetilde{\varphi}|_{r_*=0}$ with $k=\sqrt{\om^2-\widetilde{V}(r_0)}$ in (\ref{k_throat_general}). On the other hand, when allowing an additional homogeneous solution in (\ref{Phi_near_throat}), $\widetilde{\varphi}\sim e^{-ik r_*}+i(k-\om) r_*$, {we have}
\beq
{\rm Im} (I)=\om_R \left[ \left|\widetilde{\varphi}(r_*=0)
\right|^2-2  \om_I \int^{\de}_0 |\widetilde{\varphi}|^2 dr_* \right]=0,
\label{Im_I_light-like}
\eeq
{from
$({d \widetilde{\varphi}}/{dr_*})|_{r_*=0}=-i \om \widetilde{\varphi}|_{r_*=0}$ as if the solution is {\it light-like}
at the throat, similar to the case
for the black hole}
\cite{Kono:2008,Moon:2011}.
 Actually, this second case corresponds to our choice of
 vanishing Neumann boundary condition $({d \varphi}/{dr_*}|_{r_*=0}=0)$
 from the solution $\varphi= {e^{i\om r_*}\widetilde{\varphi}} \sim
 e^{i(\om-k) r_*}+i(k-\om){r_*} e^{i \om r_*}$. Now, for the simplicity
 of our discussion, let us consider only the second case
 \footnote{For the first case, the situation looks more complicated
 since, as noted in Sec. III, the solution is not light-like generally
 at the throat due to $\widetilde{V}(r_0)\neq0$ and $k_R$ is not
a simple function of $\om_R$ alone. But since the {\it sign} of $k_R$
coincides with that of $\om_R$, the argument is basically the same.},
which we have studied numerically in this paper. From
(\ref{Im_I_light-like}), one finds naively that `` non-vanishing $\om_R$ may
imply $\om_I>0$, {\it i.e.,} stable modes " \cite{Kono:2008,Moon:2011}.
However, here it is important to note that this is the only case when
$\om_R$ (or $k_R$) is ``independent" on $\om_I$: When $\om_R$ (or $k_R$) is
{\it not} independent on $\om_I$, the solution $\om_I>0$ may not be the
{\it unique} possibility,
generally. For example, if we consider $\om_R$ as a function of $\om_I$, $\om_R=\om_R (\om_I)$, (\ref{Im_I_light-like}) is generally a non-linear equation for one independent variable $\om_I$ and its solution needs not to agree with the previous naive one without separating independent variables, which may
lead to misleading solutions. This means that $\om_R \neq 0,~ \om_I<0$
(oscillating, unstable modes) can {also be} the possible solution
depending on the details of $\om_R=\om_R (\om_I)$. Actually, Fig. 14
shows the relation $\om_R \approx 1/2+ a~ \om_I +b~ \om_I^2 \cdots$ which
allows the continuous solution of $\om_R \neq 0, \om_I<0$, as implied by
the result of (\ref{omega_near_critical}) (left), as well as the usual
discontinuous solution (right, $l=0,1$ case). This
indicates the existence of more fundamental reason for the relation \footnote{In the context of {Green's function with QNMs as its} poles (see \cite{Bert:2009} for a review), it would be natural to expect the relation $\om_R=\om_R (\om_I)$ due to the Kramers-Kronig relation for the unitarity.}. The usual {result} $\om_R=0, \om_I<0$ for the {case} $l=0$ may be due to the independence of $\om_I(<0)$ from $\om_R$ after the level crossing of two initially different modes. }

Finally, in order to obtain reliable numerical
results we need to compute the partial sum with the typical truncation of the order of $N=250$ (Fig. 15). In this work, we consider up to $N=300$ for most computations, but we consider up to $N=400$ for the {level crossing} of $l=0$ QNMs in Fig. 9.

\section{Concluding Remarks}

We have studied QNMs for a massive scalar field in the background of a
{\it natural} AdS wormhole in EBI gravity, which has been recently
constructed without exotic matters. For the case where the GR limit
exists, {\it i.e.,} $\beta Q>1/2$, we have shown numerically the existence
of a BF-like bound $m_*^2<0$
{so that the} perturbation is unstable for {a
tachyonic mass} $m^2 <m_*^2$, {like} the perturbation in the global AdS.
{Furthermore, we have shown that the unstable modes ($\om_I<0$) can also have oscillatory parts ($\om_R\neq 0$) as well as non-oscillatory parts ($\om_R= 0$), depending on whether the real and imaginary parts of frequencies are dependent on each other or not, contrary to arguments in the literature.}
On the other hand, for the case where {the} GR limit does not exist,
{\it i.e.,} $\beta Q\leq 1/2$,
the BF-like bound does not exist. In this
case, the perturbation is completely ``frozen"
{above}
a certain non-tachyonic mass bound $m_c^2>0$ which is big compared to the wormhole {mass} $M$ for $\beta Q< 1/2$.

We also have shown that, for the case where the BF-like bound exists, there is a {\it level crossing}
(between the lowest pure-imaginary and complex modes) of $\om_I$ for $l=0$ and a
bouncing behavior of $\om_I$ for higher $l$. We have shown the linear
dependence of QNMs on the throat radius, {analogous to
the black hole case.}
Even though the thermodynamic implication of this behavior is not {quite} clear, it would be interesting to study its implication to the corresponding boundary field theory as in the black hole case based on the {\it AdS/CFT-correspondence} \cite{Dani:1999}. In particular, {considering higher-order contributions for small $\widetilde{r}_*$ or large $r$ regime,
our radial equation (\ref{Rad_Eq_tort_infty})}
can be approximated by a Calogero-like model \cite{Calo:1969},
\be
\left[ -\f{d^2}{d \widetilde{r}_*^2}+\f{1}{4} \widetilde{\om}^2  \widetilde{r}_*^2 +\xi \widetilde{r}_*-\f{\lambda}{\widetilde{r}_*^2}-E\right] \widetilde{\varphi} =0
\ee
{with the energy $E$}
{and} it would be interesting to study its connection to some integrable theories at the boundary.

\begin{figure}
\includegraphics[width=5.5cm,keepaspectratio]{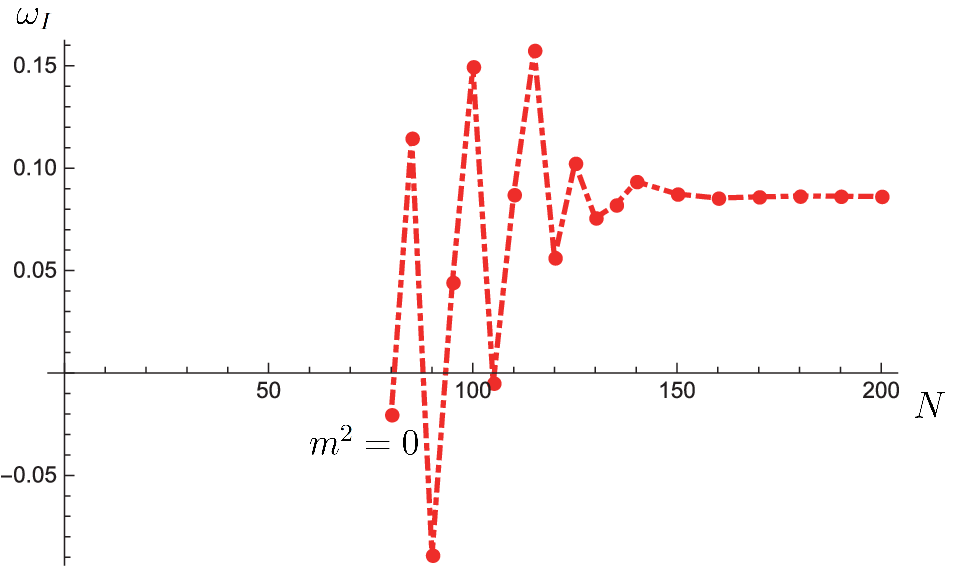}
\qquad
\includegraphics[width=5.5cm,keepaspectratio]{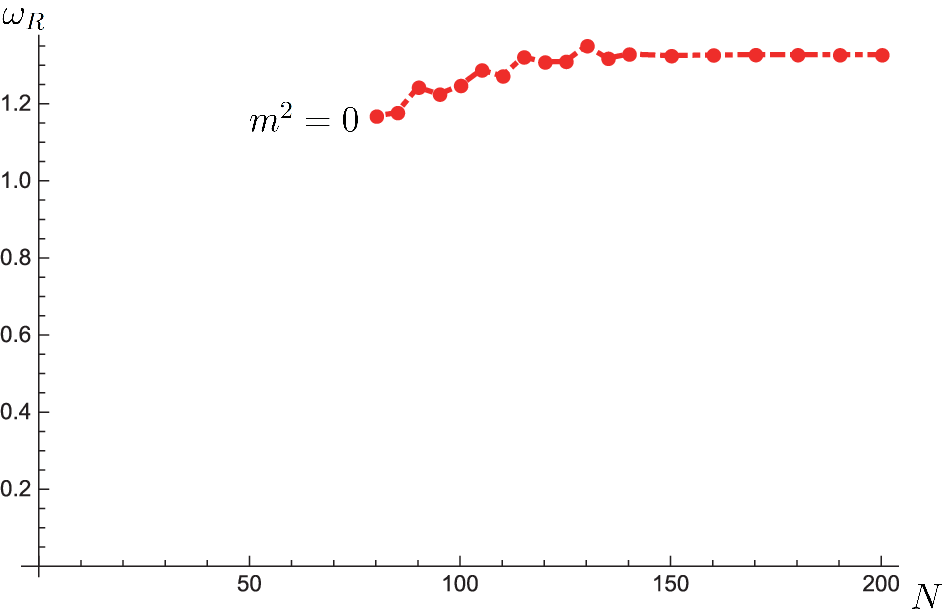}
\qquad
\includegraphics[width=5.5cm,keepaspectratio]{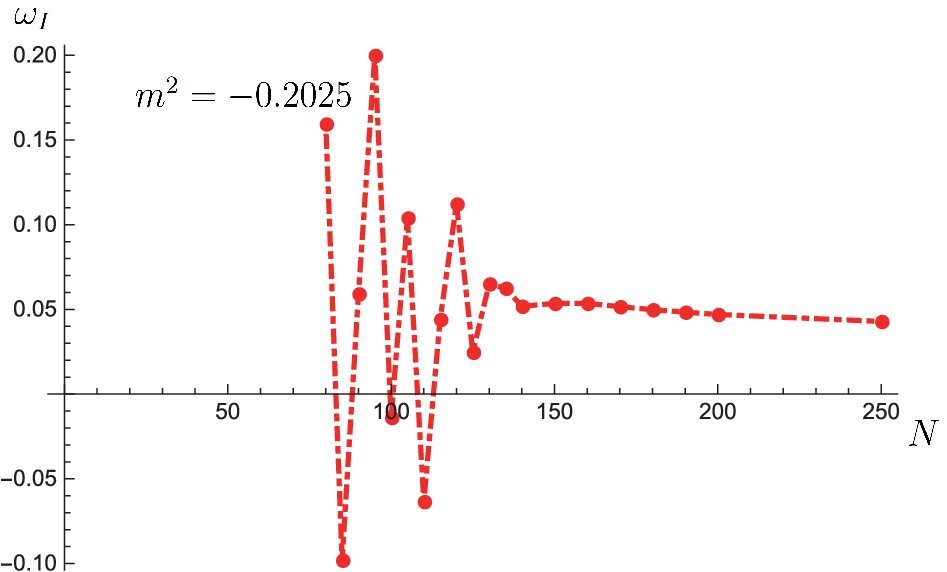}
\qquad
\includegraphics[width=5.5cm,keepaspectratio]{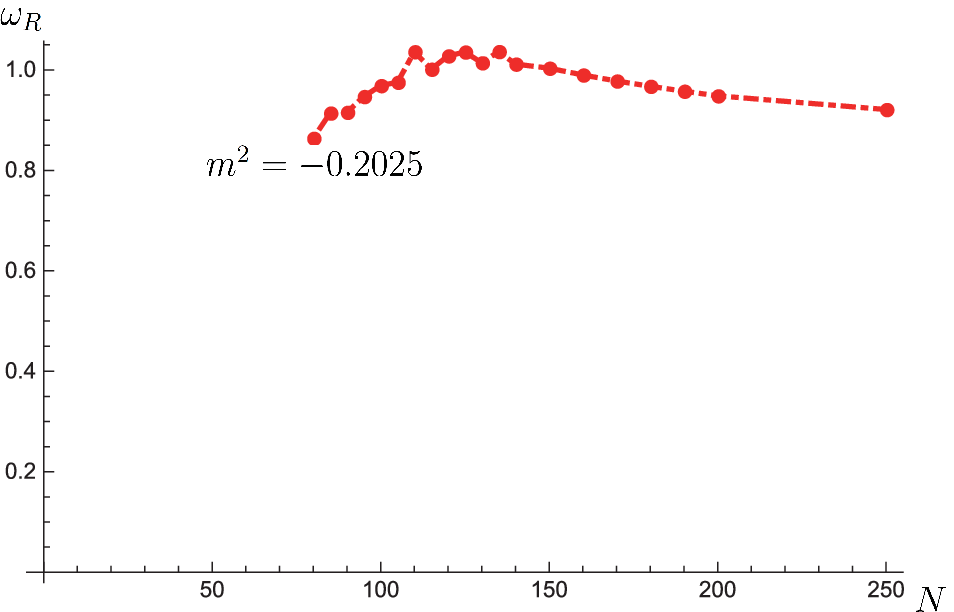}
\caption{The plots of $\om_I$ (left) and $\om_R$ (right) vs. $N$ for $m^2=0,-0.2025$ (top, bottom) with the case $M=10^{-5}$ in Fig. 6. These show the convergence of the numerical results when $N$ is about $250$. As $m^2$ becomes
{negatively larger, $N$ needs to be
higher to achieve the convergence
and these cases
are not included}
in this paper.}
\label{fig:QNM_varying_beta_vs_m}
\end{figure}

{As the final remark, it would be straightforward to extend our formalism to more general perturbations with spins, including the (gravitational, spin 2) perturbations of the wormhole space-time itself. It would be interesting to study whether ring-down phases of natural wormholes can mimic those of black holes.}
}
\section*{Acknowledgments}
{This work} was supported by {Basic Science Research
Program through the National Research Foundation of Korea (NRF) funded by
the Ministry of Education, Science and Technology (2018R1D1A1B07049451 (COL),
2016R1A2B401304 (MIP)).}

\newcommand{\J}[4]{#1 {\bf #2} #3 (#4)}
\newcommand{\andJ}[3]{{\bf #1} (#2) #3}
\newcommand{\AP}{Ann. Phys. (N.Y.)}
\newcommand{\MPL}{Mod. Phys. Lett.}
\newcommand{\NP}{Nucl. Phys.}
\newcommand{\PL}{Phys. Lett.}
\newcommand{\PR}{Phys. Rev. D}
\newcommand{\PRL}{Phys. Rev. Lett.}
\newcommand{\PTP}{Prog. Theor. Phys.}
\newcommand{\hep}[1]{ hep-th/{#1}}
\newcommand{\hepp}[1]{ hep-ph/{#1}}
\newcommand{\hepg}[1]{ gr-qc/{#1}}
\newcommand{\bi}{ \bibitem}

\end{document}